\documentclass[twocolumn]{revtex4-1}
\usepackage[utf8]{inputenc}
\usepackage{graphicx}
\usepackage{subfigure}
\usepackage{braket}
\usepackage{amsmath}
\usepackage{hyperref}

\begin{document}
\title{Predicting novel superconducting hydrides using machine learning approaches}

\author{Michael J. Hutcheon}
\email{mjh261@cam.ac.uk}
\affiliation
{
    Theory of Condensed Matter Group,
    Cavendish Laboratory,
    J.~J.~Thomson Avenue,
    Cambridge CB3 0HE,
    United Kingdom
}

\author{Alice M. Shipley}
\email{ams277@cam.ac.uk}
\affiliation
{
    Theory of Condensed Matter Group,
    Cavendish Laboratory,
    J.~J.~Thomson Avenue,
    Cambridge CB3 0HE,
    United Kingdom
}

\author{Richard J. Needs}
%\email{rn11@cam.ac.uk}
\affiliation
{
    Theory of Condensed Matter Group,
    Cavendish Laboratory,
    J.~J.~Thomson Avenue,
    Cambridge CB3 0HE,
    United Kingdom
}

\date{\today}

\begin{abstract}
The search for superconducting hydrides has, so far, largely focused on finding materials exhibiting the highest possible critical temperatures ($T_c$). This has led to a bias towards materials stabilised at very high pressures, which introduces a number of technical difficulties in experiment. Here we apply machine learning methods in an effort to identify superconducting hydrides which can operate closer to ambient conditions. The output of these models informs subsequent crystal structure searches, from which we identify stable metallic candidates prior to performing electron-phonon calculations to obtain $T_c$. Hydrides of alkali and alkaline earth metals are identified as especially promising; of particular note, a $T_c$ of up to 115\ K is calculated for RbH$_{12}$ at 50\ GPa, which extends the operational pressure-temperature range occupied by hydride superconductors towards ambient conditions.
\end{abstract}

\maketitle

\section{Introduction}
While hydrogen is predicted to be a room-temperature superconductor at very high pressures \cite{ashcroft1968}, metal hydrides, in which the hydrogen atoms are ``chemically pre-compressed", are predicted to exhibit similar behaviour in experimentally-accessible regimes \cite{Gilman1971546, ashcroft2004}. In recent years, potential superconductivity has been investigated in many compressed hydrides, including scandium \cite{durajski2014}, sulfur \cite{duan2014, drozdov2015, errea2015}, yttrium \cite{kim2009, li2015, liu2017,peng2017,heil2019,troyan2019,kong2019, shipley2020}, calcium \cite{wang2012}, actinium \cite{semenok2018}, thorium \cite{kvashnin2018}, pnictogen \cite{fu2016}, praseodymium \cite{zhou2019}, cerium \cite{salke2019,li2019}, neodymium \cite{zhou2019neodymium}, lanthanum \cite{liu2017,peng2017,somayazulu2019, drozdov2018v2, shipley2020,kruglov2020} and iron hydrides \cite{majumdar2017, kvashnin2018iron, heil2018}. Several reviews summarising recent developments in the field are available \cite{duan2017review, zurek2019, flores2019, boeri2019, pickard2019, hydrides_review_1}. Inspired by known superconductors, researchers have also attempted to increase $T_c$ by chemical means; replacing atoms in known structures and assessing stability and superconductivity \cite{chang2019}, doping known binaries with more electronegative elements to make ternary hydrides \cite{sun2019}, and mapping alchemical phase diagrams \cite{heil2015}.

Experimental measurements of superconductivity in high-pressure hydrides have helped to address several misconceptions about conventional superconductivity, fuelling hope that it may be achieved at ambient temperature and waving a definitive farewell to the Cohen-Anderson limit \cite{cohen1972}. The associated theoretical studies have demonstrated that the crystal structures and superconducting properties of real materials can now be accurately predicted from first principles. 

In this work, we train machine learning models on a set of literature data for superconducting binary hydrides. Machine learning has previously been used in modelling hydride superconductors, with a focus on predicting the maximum obtainable critical temperature for a given composition \cite{semenok2018distribution}. However, on examination of the literature (see Fig.\ \ref{fig:hydrides_pt}), it becomes apparent that the pursuit of superconductivity close to ambient conditions is as much about reducing the required pressure as it is about increasing the critical temperature. This is especially important given that working at high pressure can often present a far greater experimental challenge than working at low temperature. In this work, we therefore model critical temperature and operational pressure on an equal footing. Our models are used to inform the choice of composition for crystal structure searches and subsequent electron-phonon calculations, with the aim of extending the operation of hydride superconductors towards ambient conditions.

\section{Trends in Hydrides}
\label{sec:trends_in_hydrides}

\begin{figure*}
    \centering
    \includegraphics[width=\textwidth]{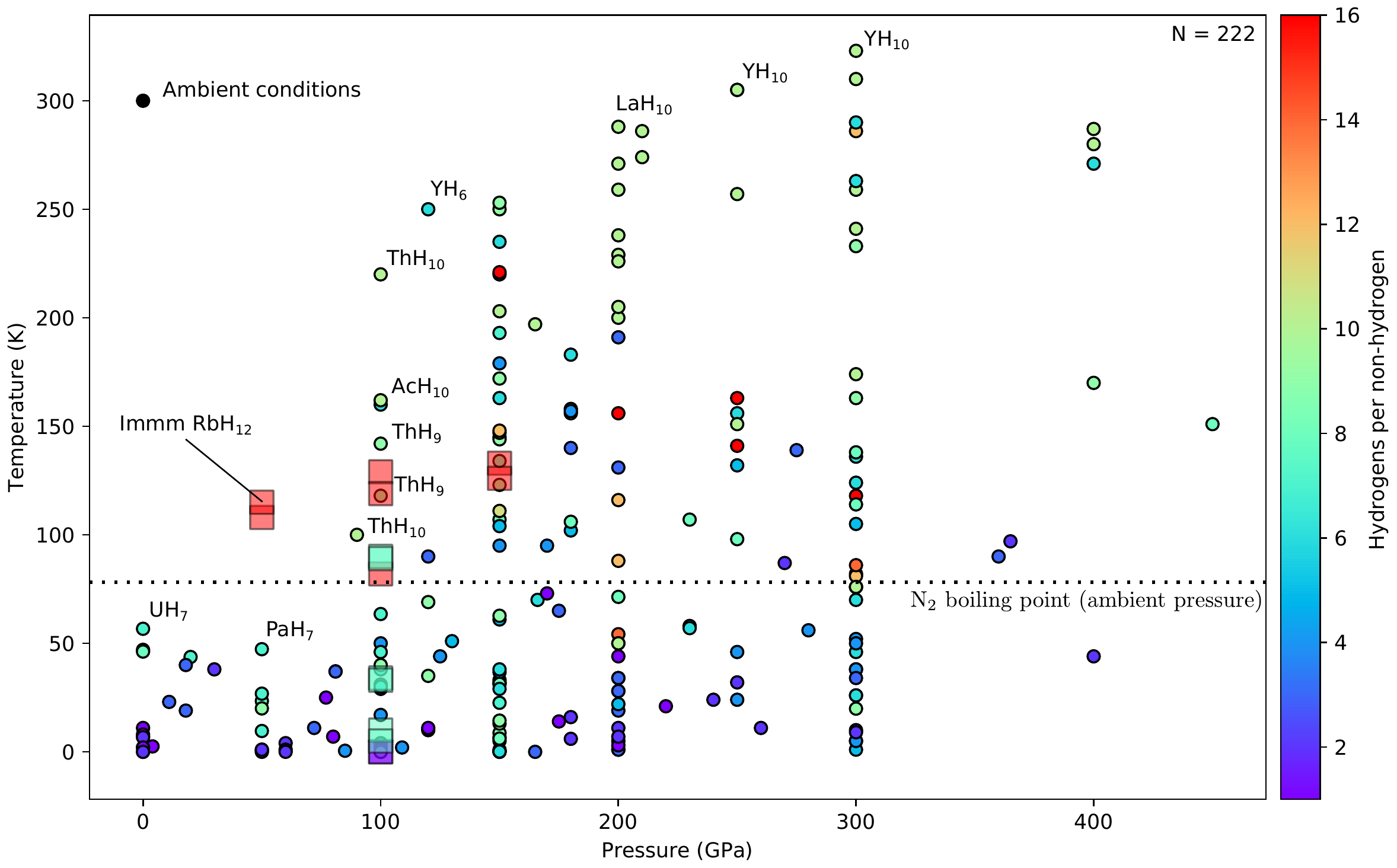}
    \caption{The critical temperatures of binary hydrides at various pressures found in the literature are shown as circles. Materials on the frontier towards ambient conditions are labeled. Multiple points with the same stoichiometry arise from $T_c$ for a particular phase calculated at different pressures, or from different structural phases of the same material. New structures found in this work, with $T_c$ calculated using DFPT (as reported in Table\ \ref{tab:new_materials}), are shown as translucent squares; of note is $Immm$-RbH$_{12}$ (labeled, see also Fig.\ \ref{fig:hydride_structures}), which extends the frontier significantly.}
    \label{fig:hydrides_pt}
\end{figure*}

A large amount of computational - and some experimental - data for the binary hydrides is available in the literature \cite{geballe2018, troyan2019, errea2019, heil2019, kong2019, Xie:od3007, PhysRevB.86.014118, doi:10.1063/1.4898145, PhysRevB.87.054107, TSE20091944, semenok2018, kvashnin2018, wang2012, zemla2019, kruglov2018uranium, C5RA11459D, gu2017high, durajski2014, esfahani2017superconductivity, C6RA11862C, zhou2019, Liu_2017, semenok2019superconductivity, peng2017, liu2017, zarifi2018crystal, Shanavas2016, C6CP08036G, kresin2018high, semenok2018distribution, Liu2015, OHLENDORF1979721, doi:10.1063/1.4866179, PhysRevB.88.184104, Zhuang2017, Yu2015, Spitsyn1982, C6RA25591D, kvashnin2018iron, C5CP06617D, Liu2015, ye2018high, Skoskiewicz_1974, li2017superconductivity, PhysRevB.89.064302, tanaka2017, PhysRevLett.110.165504, doi:10.1063/1.4821287, C4RA14990D, li2016crystal, Szcze_niak_2013, hooper2013, Liu2015, Yao_2007, zhuang2017pressure, Eremets1506, Li15708, Jin9969, liu2018, drozdov2018, somayazulu2019, drozdov2018v2, li2015} (values from these references form our dataset, shown in Fig.\ \ref{fig:hydrides_pt}). In some subsets of hydrides certain material properties show a simple dependence on the properties of the non-hydrogen element. For example, in the alkaline earth hydrides the van der Waals radius of the ion is well correlated with the metallization pressure \cite{alkaline_hydrides_radius_trend}. However, obtaining strong electron-phonon coupling at low pressures is, in general, a more complicated process; simple correlations between composition and operational pressure or critical temperature are therefore absent in the dataset as a whole. We look at more complicated trends by constructing machine learning models of critical temperature and operational pressure which take as input a set of easily-obtained material descriptors. For a particular element E and corresponding binary hydride EH$_n$ these descriptors are
\begin{itemize}
    \setlength\itemsep{-0.5em}
    \item hydrogen content ($n$)
    \item van der Waals radius of E
    \item atomic number of E
    \item mass number of E
    \item numbers of $s$, $p$, $d$ and $f$ electrons in the (atomic) electron configuration of E
\end{itemize}
Once constructed, we apply the model to all materials with the chemical composition EH$_{n}$, where E is any element in the periodic table and $n \in [1,2,\dots,32)$ \footnote{A maximum of 31 hydrogens per atom was chosen to avoid over-extrapolation from the dataset (where the maximum is 16).}. From these, the materials which are predicted to exhibit superconductivity closest to ambient conditions serve as a guide for searches for new binary hydrides.

\subsection{Neural network}
\begin{figure} 
    \centering
    \includegraphics[width=\columnwidth]{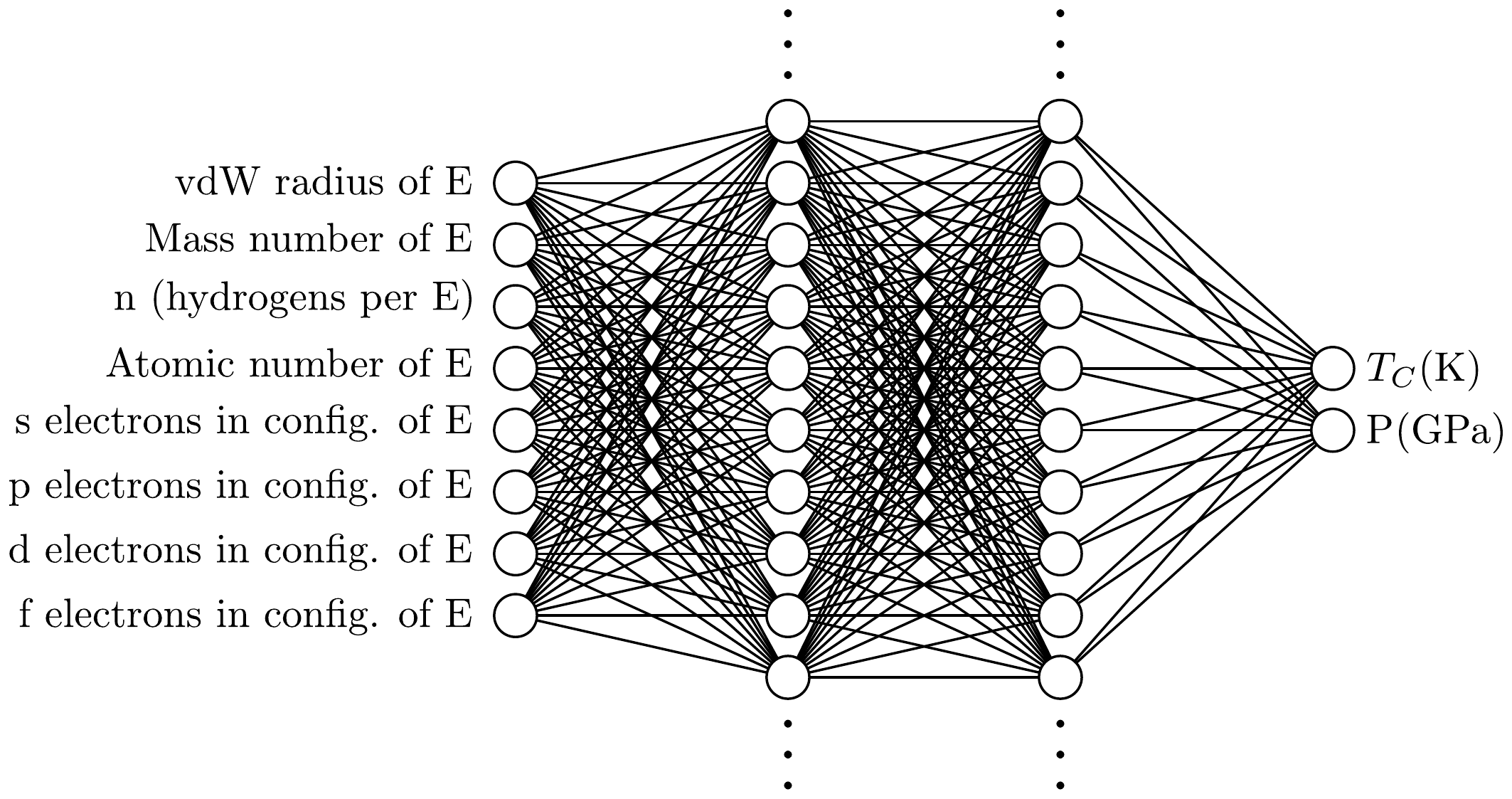}
    \caption{Topology of our neural network model. An input layer is fed to the material descriptors for the hydride EH$_{n}$, one per input node. This layer then feeds two densely-connected intermediate layers (of 32 nodes each), the last of which feeds the output layer with one temperature node and one pressure node.}
    \label{fig:neural_network}
\end{figure}

We train a fully-connected neural network (using the Keras frontend to the Tensorflow machine-learning library \cite{keras,tensorflow}), with the topology shown in Fig.\ \ref{fig:neural_network}, on the dataset shown in Fig.\ \ref{fig:hydrides_pt}. The squared absolute error $|(\Delta T_c, \Delta P)|^2$ between the predicted and literature values serves as our cost function, which we minimize using the \textit{Adam} stochastic optimizer \cite{adam_optimizer}. The input (and expected output) data is positive definite and therefore has a non-zero mean and is not normally distributed, prompting the use of self-normalizing activation functions \cite{self_normalizing_neural_nets,exponential_linear_unit_activation} to improve training behaviour. Since the number of data points is comparable to the number of parameters in our network, the risk of over-fitting becomes significant. To mitigate this, we split the data into a randomly selected validation set (consisting of 25\% of the initial data points) and a training set (consisting of the other 75\%). Once the model starts over-fitting to the training data the validation set error starts to increase, allowing us to choose the model parameters from the training epoch for which the validation set error is minimal. We cross-validate the results by repeating this process 64 times and averaging the predictions - this is an approximation of leave-$p$-out cross-validation with $p=25\%$ of the dataset. We also apply $L_2$ regularization to the parameters in the intermediate dense nodes to decrease the propensity for over-fitting, improving the convergence of this cross-validation scheme.

\begin{figure}
    \centering
    \includegraphics[width=\columnwidth]{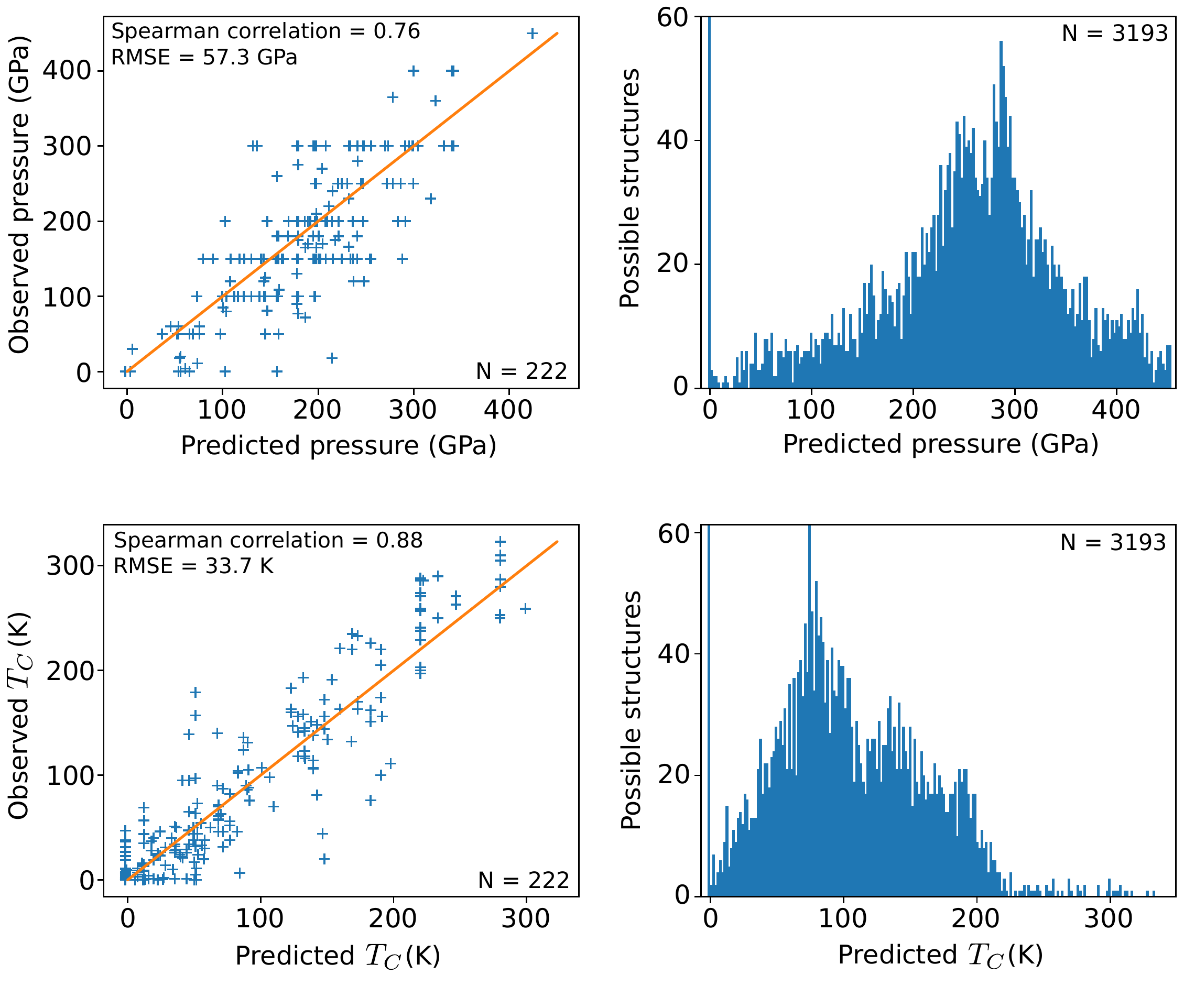}
    \caption{Behaviour of our machine learning model of critical temperatures and associated operational pressures for binary hydrides. The correlation between the predicted and observed values for the data in the literature is shown, as well as the resulting distribution of pressures and temperatures when the model is applied to the set of all possible binary hydrides as defined in Section\ \ref{sec:trends_in_hydrides}.}
    \label{fig:ml_hydrides}
\end{figure}

\subsection{Model behaviour}
\begin{figure}
    \centering
    \includegraphics[width=\columnwidth]{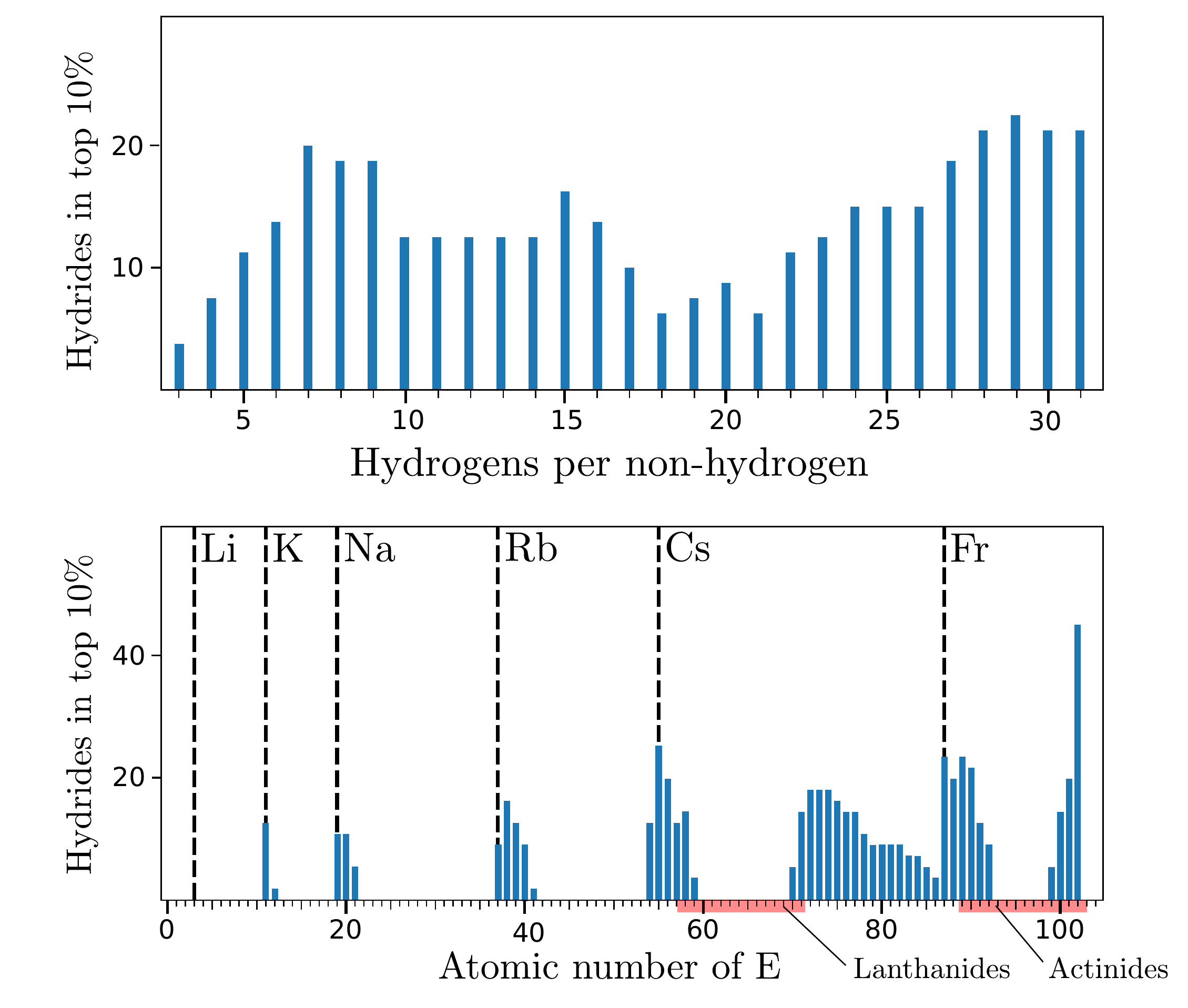}
    \caption{Distribution of hydrogen atoms per non-hydrogen atom and atomic number of the non-hydrogen element for the 10\% of hydrides that our machine learning model predicted to exhibit superconductivity closest to ambient conditions (i.e., the 10\% with lowest $D$). Black dashed lines indicate the atomic numbers of alkali metals.}
    \label{fig:ml_best_feature_dist}
\end{figure}

\begin{figure*}
    \centering
    \includegraphics[width=\textwidth]{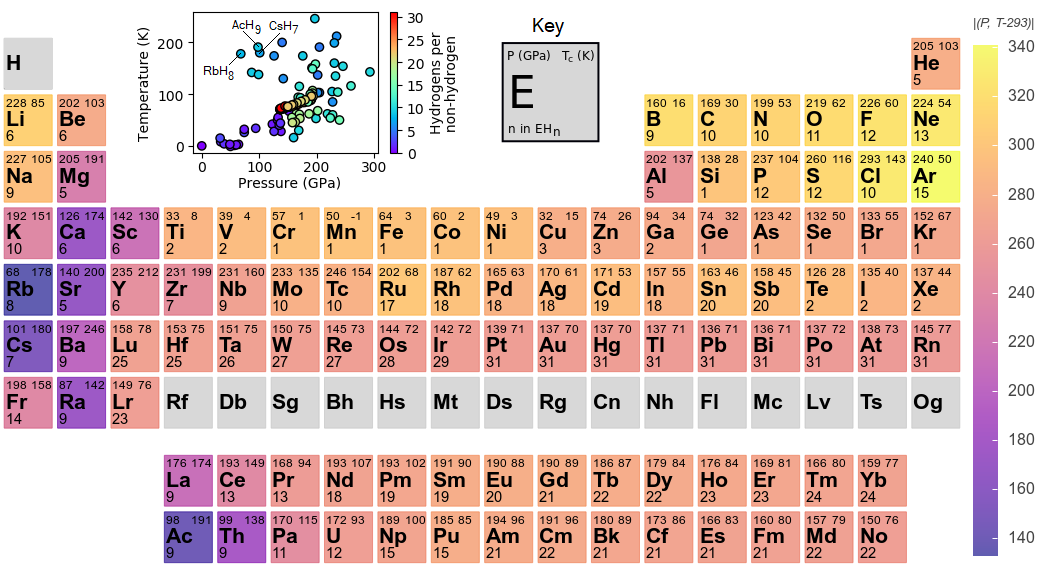}
    \caption{The periodic table of optimal binary hydrides according to our machine learning model. The predicted critical temperature, corresponding pressure and optimal hydrogen content is shown for each element. Elements are colored according to the predicted distance from ambient-condition superconductivity $D=|(P, T_c-293)|$. Inset: the distribution in pressure-temperature space of these predictions. We note that we did not explicitly prevent the neural network from predicting negative critical temperatures and it does so for MnH. However, in general, the machine learning model has learned that critical temperatures should be positive (see the lower right panel of Fig.\ \ref{fig:ml_hydrides}).}
    \label{fig:ml_periodic_table}
\end{figure*}

The basic behaviour of the machine learning model is shown in Fig.\ \ref{fig:ml_hydrides}. We see that it achieves reasonable correlation with the literature values and predicts sensible pressures and temperatures for unseen materials. To gain insight into properties which favour ambient-condition superconductivity we define a measure of distance $D=|(P, T_c-293)|_1$. This distance decreases as we move towards ambient conditions from the pressure-temperature region containing the known hydrides (see Fig.\ \ref{fig:hydrides_pt}). In Fig.\ \ref{fig:ml_best_feature_dist} we plot the distribution of material properties for the 10\% of hydrides predicted to exhibit superconductivity closest to ambient conditions (i.e., the 10\% with lowest $D$). We can see that the model predicts the heavy alkali and alkaline earth metal hydrides to be the best candidates, with the number of close-to-ambient materials then decreasing as we go across each period. The distribution of the number of hydrogen atoms is more uniform, suggesting it is necessary to consider a range of different stoichiometries for each composition. These conclusions are reinforced by the construction of a simple linear regression model \cite{supplement}, which reproduces the general trends exhibited by the machine learning model (but, unsurprisingly, exhibits worse correlation with the literature values). The predicted optimal (minimum $D$) hydride compositions from the machine learning model are shown for each element of the periodic table in Fig.\ \ref{fig:ml_periodic_table}.

We note that the points included in our dataset will be of varying quality, come from different research groups, and are of both experimental and theoretical origin. The majority are theoretical and calculated within the harmonic approximation. Although it has been shown that anharmonicity can affect the calculated critical temperature for hydrides \cite{errea2013,errea2015}, there is insufficient data in the literature to build a model exclusively from anharmonic results. However, since we only seek to extract general trends, which will serve simply to inform areas of focus for structure searching, the dataset is sufficient for our purposes.

\section{Structure searching}
The models constructed in the previous section point towards the alkali and alkaline earth metal hydrides as some of the best candidates for superconductivity near ambient conditions. From these, we studied caesium and rubidium hydrides; these systems were chosen due to their predicted proximity to superconductivity at ambient conditions (see Figs.\ \ref{fig:ml_best_feature_dist} and \ref{fig:ml_periodic_table}) and the fact that they have not been studied extensively in the past, unlike the hydrides of other elements in these two groups. Caesium and rubidium polyhydrides have been studied previously using structure searching methods in Refs.\ \cite{shamp2012} and \cite{hooper2012}, respectively, although potential superconductivity was not investigated in either case.

Our structure searching calculations were performed using \textit{ab initio} random structure searching (AIRSS) \cite{pickard2011, needs2016} and the plane-wave pseudopotential code \textsc{castep} \cite{castep2005}. Since our models suggest that a wide range of stoichiometries should be considered, convex hulls were constructed using AIRSS and qhull \cite{barber1996} in order to identify those which are stable at 50, 100 and 200\ GPa \cite{supplement}. The Perdew-Burke-Ernzerhof (PBE) generalised gradient approximation \cite{pbe1996}, \textsc{castep} QC5 pseudopotentials, a 400\ eV plane-wave cut-off and a $\mathbf{k}$-point spacing of $2\pi\times$0.05\ \r{A}$^{-1}$ were used in all searches. The Cs-H convex hulls calculated in this work at 100 and 200\ GPa both partially agree with the hull calculated at 150\ GPa in Ref.\ \cite{shamp2012}. Once stable stoichiometries had been identified, additional AIRSS searches for RbH$_3$, RbH$_5$, RbH$_9$, RbH$_{11}$, RbH$_{12}$, CsH$_5$, CsH$_7$, CsH$_{13}$ and CsH$_{15}$ using the same parameters and pseudopotentials were performed at 100 and 200\ GPa.

\section{Selecting candidate structures}
For each selected stoichiometry, the enthalpy was calculated as a function of pressure for the most stable structures arising from the AIRSS search. These geometry optimisations were performed using \textsc{quantum espresso} \cite{QE-2009,QE-2017}, the PBE functional, a 950\ eV cut-off, ultrasoft pseudopotentials \cite{supplement} and a $\mathbf{k}$-point spacing of $2\pi\times$0.02\ \r{A}$^{-1}$. The electronic density of states (DOS) at the Fermi energy was also evaluated for each structure at 50\ GPa and 150\ GPa in order to identify metallic structures. We were then able to limit our interest to structures which were both energetically competitive (according to the enthalpy plots) and had a considerable DOS at the Fermi energy in the low-pressure region (25-125\ GPa). Full lists of the competitive structures predicted here are available, along with the enthalpy plots and DOS values, in the supplementary material \cite{supplement}. The remaining candidates, for which electron-phonon coupling calculations were performed, include $C2/m$-RbH$_{12}$, $Immm$-RbH$_{12}$, and various CsH$_7$ and RbH$_3$ structures (see Table\ \ref{tab:new_materials}).

\section{Electron-phonon coupling and superconductivity}
The Hamiltonian of a coupled electron-phonon system is given by
\begin{multline}
H=\sum_{kn}\epsilon_{nk}c_{nk}^\dagger c_{nk}+\sum_{q\nu}\omega_{q\nu}(a_{q\nu}^\dagger a_{q\nu}+\frac{1}{2}) +\\ \frac{1}{\sqrt{N_p}}\sum_{kqmn\nu}g_{mn\nu}(k,q)c_{m,k+q}^\dagger c_{nk}(a_{q\nu}+a_{-q\nu}^\dagger)
\label{elph_H}
\end{multline}
In this work, we calculate the electronic Kohn-Sham eigenvalues $\epsilon_{nk}$, phonon frequencies $\omega_{q,\nu}$, and electron-phonon coupling constants $g_{mn\nu}(k,q)$ appearing in $H$ from first-principles using density functional perturbation theory (DFPT) as implemented in the \textsc{quantum espresso} code \cite{QE-2009,QE-2017}. The resulting Hamiltonian is then treated within Migdal-Eliashberg theory \cite{migdal1958, eliashberg1960, elk_eliashberg} where we solve the Eliashberg equations using the \textsc{elk} code \cite{elk_code}. This gives us the superconducting gap as a function of temperature, from which we obtain a prediction for $T_c$.

To carry out these calculations, we use the PBE functional, the same ultrasoft pseudopotentials as in the geometry optimisations, an 820\ eV plane-wave cut-off, and a $\mathbf{q}$-point grid with a spacing of $\approx 2\pi\times 0.1$ \r{A}$^{-1}$ (e.g., a $2\times2\times2$ grid for a 26-atom unit cell of RbH$_{12}$). Two separate $\mathbf{k}$-point grids are used (of $6^3$ and $8^3$ times the size of the $\mathbf{q}$-point grid, respectively), allowing us to determine the optimal double-delta smearing width necessary to calculate the critical temperature \cite{wierzbowska2005, shipley2020}. 

\begin{figure}
\centering
\includegraphics[width=\columnwidth]{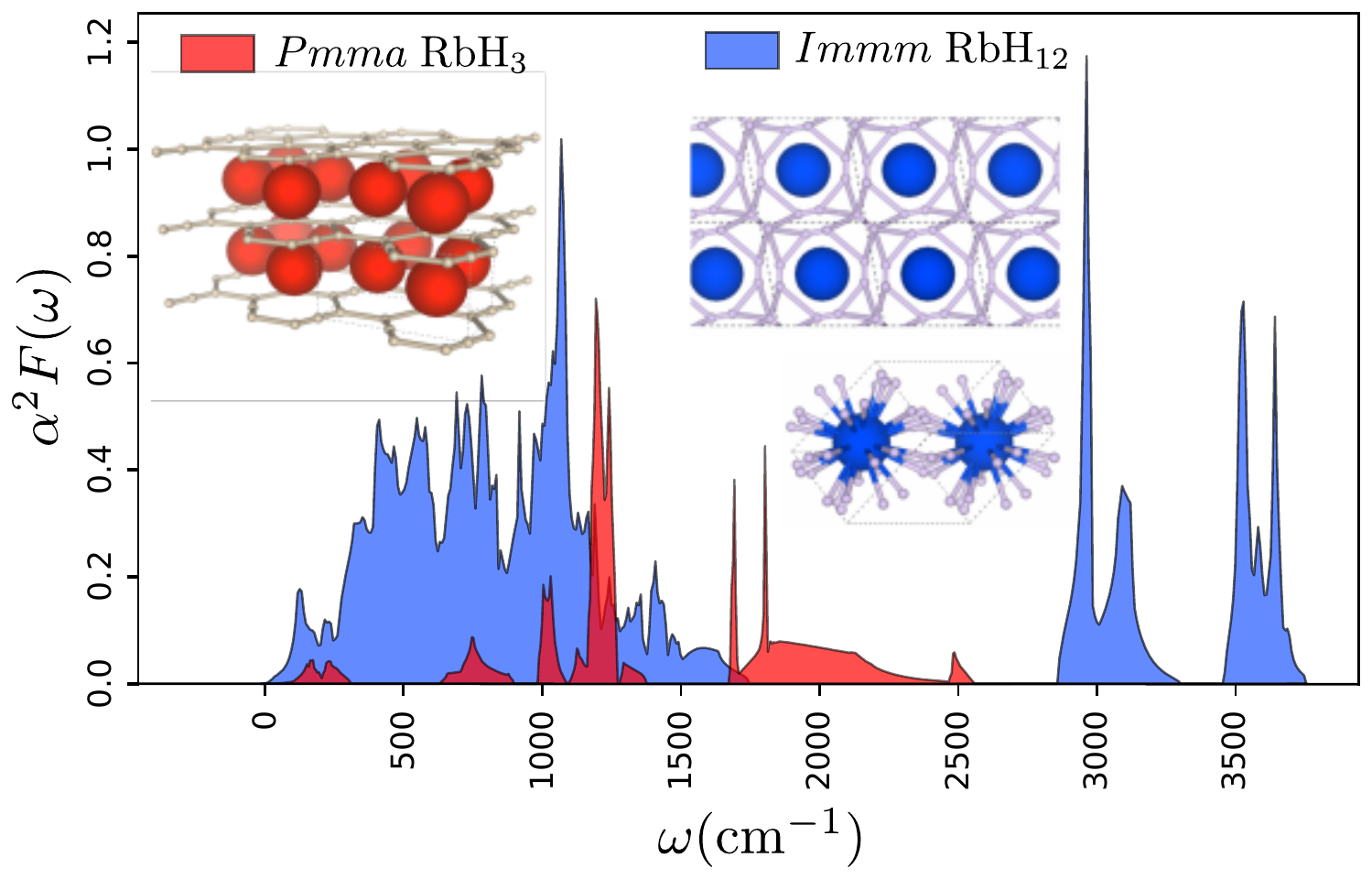}
\caption{The Eliashberg function for $Immm$-RbH$_{12}$ and $Pmma$-RbH$_3$ found in our AIRSS searches (see Table\ \ref{tab:new_materials}). It is clear to see the enhanced high-frequency part of the Eliashberg function for Immm-RbH$_{12}$, arising from the hydrogen cage. In contrast, the Eliashberg function for the layered RbH$_3$ structure does not extend to such high frequencies. This effect can also be seen from the phonon linewidths (plotted along with the phonon dispersion in the supplementary material \cite{supplement}).}
\label{fig:hydride_structures}
\end{figure}

Electron-phonon calculations were performed for a range of competitive RbH$_{12}$, CsH$_7$ and RbH$_3$ structures predicted in this work and the results are shown in Table\ \ref{tab:new_materials}. The highest-$T_c$ results arise from structures with a cage-like arrangement of hydrogen atoms surrounding a central non-hydrogen element. The electronic states that originate from these cages are near the Fermi level, and are strongly coupled together by cage vibrations. This provides the phonon-mediated pairing mechanism necessary for conventional superconductivity. Combined with a high average phonon frequency, owing to the light mass of the hydrogen atoms, this results in a high critical temperature (c.f the Allen-Dynes equation \cite{allen1975}). This can be seen directly by looking at the Eliashberg function, shown in Fig.\ \ref{fig:hydride_structures}, for two illustrative structures from Table\ \ref{tab:new_materials}. The enhanced high-frequency portion of the Eliashberg function for the high-$T_c$ cage-like RbH$_{12}$ structure is apparent. In contrast, strong electron-phonon coupling is absent at high phonon frequencies for states near the Fermi level in the layered RbH$_3$ structure, leading to a negligible $T_c$. It is perhaps unsurprising that our machine learning model suggests such compositions, despite their resulting unfavorable structures, as it is trained on mostly cage-like structures. As a result, the model may implicitly assume that compositions it is given will behave as if they adopt cage-like arrangements, leading to an overestimation of $T_c$. Despite this, most of the structures found are high-$T_c$ cage-like superconductors, of which $Immm$-RbH$_{12}$ is particularly interesting due to its location in Fig.\ \ref{fig:hydrides_pt}.

Supplementing structure searching techniques with predictions from machine learning has allowed us to target novel regions of pressure-temperature space. We have therefore been able to identify low-pressure hydride superconductors without having to perform a large number of expensive electron-phonon calculations. It can be seen from Fig.\ \ref{fig:hydrides_pt} that the hydrides predicted in this work are biased towards ambient conditions when compared to the dataset as a whole.

\begin{table}
\centering
\begin{tabular}{llll}
    Stoichiometry & Space group & Pressure (GPa) & $T_c$ (K)  \\
    \hline \hline
    RbH$_{12}$   & $C2/m$      & 50             & 108        \\
    RbH$_{12}$   & $C2/m$      & 100            & 129        \\
    RbH$_{12}$   & $C2/m$      & 150            & 133        \\
    RbH$_{12}$   & $Cmcm$      & 100            & 82         \\
    RbH$_{12}$   & $Immm$      & 50             & 115        \\
    RbH$_{12}$   & $Immm$      & 100            & 119        \\
    RbH$_{12}$   & $Immm$      & 150            & 126        \\
    \hline
    CsH$_{7}$    & $P1$        & 100            & 90         \\
    CsH$_{7}$    & $I4mm$      & 100            & 34         \\
    CsH$_{7}$    & $P4mm$      & 100            & 33         \\
    CsH$_{7}$    & $I4/mmm$     & 100            & 10         \\
    CsH$_{7}$    & $Cm$        & 100            & 5          \\
    CsH$_{7}$    & $Cmc2_1$     & 100            & 89         \\
    \hline
    RbH$_{3}$    & $Pmma$      & 100            & 0          \\
    RbH$_{3}$    & $Cmmm$      & 100            & 0          \\
\end{tabular}
\caption{Critical temperatures calculated using DFPT for promising hydride compositions - the structures listed here were found in this work using AIRSS and are available in an online repository \cite{dataset}. The data in this Table is also shown in Fig.\ \ref{fig:hydrides_pt} for comparison with previous results in the literature.}
\label{tab:new_materials}
\end{table}

\section{Testing potential screening techniques for high-\texorpdfstring{$T_c$}{Tc} candidates}
In this work we also tested two potential methods for cheaply estimating $T_c$ ordering between structures. Good superconductivity in hydrides generally requires hydrogenic states close to the Fermi level, which (as exemplified by the findings of this work) often means favouring cage-like structures and avoiding structures with molecular-character H$_2$ units. It is therefore possible that the hydrogen-derived DOS normalised by the total DOS at the Fermi energy, $N_H(E_F)/N(E_F)$, may give some indication of whether a particular structure will exhibit high-$T_c$ superconductivity. Here we also consider the hydrogen-derived electron-phonon coupling estimates ($\eta_H$) from Gaspari-Gyorffy theory \cite{gaspari1972} and test whether these two quantities could provide a method for ranking different structures (of the same stoichiometry and at the same pressure) before performing expensive electron-phonon calculations. We implemented Gaspari-Gyorffy theory within the \textsc{elk} code \cite{elk_code}. The basics of this theory and its use here are explained in Appendix \ref{GGtheory_section}. 

The calculated $T_c$ values for the structures predicted and studied in this work allowed us to directly assess these potential screening methods. We observe that $\eta_H$ correctly predicts the $T_c$ ordering for the RbH$_{12}$ structures at fixed pressure, as was the case for the LaH$_{10}$ and YH$_{10}$ systems on which preliminary tests were performed \cite{supplement}. The two quantities tested here often predict the same general trends, but the DOS ratio is cheaper to calculate since it can be obtained using a pseudopotential code. Unfortunately, $N_H(E_F)/N(E_F)$ appears to be much less predictive for the CsH$_7$ structures and the performance of $\eta_H$ is also mixed \cite{supplement}. The use of these quantities for screening applications therefore requires further investigation and testing in a wider variety of systems.

\section{Conclusions}
Having identified the need to reduce the operational pressure of hydride superconductors, we searched for crystal structures which would exhibit superconductivity in novel regions of pressure-temperature space. We found that guiding structure searching techniques using a machine learning model allowed us to target the most promising regions. Specifically, we constructed models of critical temperature and operational pressure trained on the available theoretical and experimental results for binary hydride superconductors. Several novel systems were identified as promising superconductors closer to ambient conditions; here we focused on Cs and Rb hydrides, using AIRSS to identify stable stoichiometries and predict crystal structures. Other promising candidates included Ca, Sr, Ba, Ra, Ac, Th, La and Sc hydrides, most of which had already been theoretically studied to some extent \cite{wang2012, tanaka2017, hydrides_review_1, hooper2013, semenok2018, kvashnin2018, semenok2019superconductivity, liu2017, peng2017, liu2018, kruglov2020, durajski2014}. Critical temperatures of energetically-competitive candidate structures were then calculated from first principles using DFPT. A $T_c$ of up to 115\ K was calculated for RbH$_{12}$ at 50\ GPa, which represents a significant extension towards ambient-condition superconductivity from our dataset.

\section*{Acknowledgements}
We thank Po-Hao Chang for useful discussions regarding Gaspari-Gyorffy theory. M.J.H.\ acknowledges the EPSRC Centre for Doctoral Training in Computational Methods for Materials Science for funding under grant number EP/L015552/1. A.M.S.\ acknowledges funding through an EPSRC studentship. R.J.N.\ is supported by EPSRC under Critical Mass Grant EP/P034616/1 and the UKCP consortium grant EP/P022596/1. We are grateful for computational support from the UK national high performance computing service, ARCHER, for which access was obtained via the UKCP consortium and funded by EPSRC grant ref EP/P022561/1. This work was also performed using resources provided by the Cambridge Service for Data Driven Discovery (CSD3) operated by the University of Cambridge Research Computing Service ( \url{www.csd3.cam.ac.uk}), provided by Dell EMC and Intel using Tier-2 funding from the EPSRC (capital grant EP/P020259/1), and DiRAC funding from the STFC (\url{www.dirac.ac.uk}).

\appendix
\section{Gaspari-Gyorffy theory}
\label{GGtheory_section}
McMillan \cite{mcmillan1968} showed that for strong-coupled superconductors the electron-phonon coupling constant, $\lambda$, can be expressed as
\begin{equation}
\label{couplconst}
\lambda=2\int\frac{d\omega\alpha^2(\omega)F(\omega)}{\omega}=\frac{N(E_F)\braket{I^2}}{M\braket{\omega^2}}
\end{equation}
$\lambda$ can also be reformatted as
\begin{equation*}
\lambda=\frac{\eta}{M\braket{\omega^2}}
\end{equation*}
where $\eta$ is the so-called Hopfield parameter. Hopfield was one of the first to stress the importance of the local environment in determining $\lambda$ \cite{hopfield1969}. In situations where we have nearly perfect separation of vibrational modes into those of different atomic character (such as we may see in hydrides) we can write
\begin{equation}
\label{totlambda}
\lambda=\sum_j\lambda_j=\sum_j\frac{\eta_j}{M_j\braket{\omega_j^2}}
\end{equation}
where $j$ is the atom type.
\newline
\newline The quantity $\braket{I^2}$ appearing in Eq.\ \ref{couplconst} can be approximated using Gaspari-Gyorffy (GG) theory \cite{gaspari1972}. Recent work has emerged using this theory for metal hydrides under high pressure \cite{papaconstantopoulos2015, chang2019} despite it originally being designed for elemental transition metals. The theory, based on the rigid muffin-tin approximation (RMTA), relies on several approximations \cite{papaconstantopoulos2015} and allows us to reformulate the electron-phonon interaction in terms of phase shifts for a scattering potential. A self-consistent DOS calculation is thus all that is required to calculate $\braket{I^2}$ for each atom type and hence obtain $\eta_j$. The GG equation is
\begin{multline}
\braket{I^2}=\frac{E_F}{\pi^2N^2(E_F)}\\\sum_l\frac{2(l+1)\sin^2(\delta_{l+1}-\delta_l)N_l(E_F)N_{l+1}(E_F)}{N_l^{(1)}N_{l+1}^{(1)}}
\end{multline}
where $N_l^{(1)}$ is the free-scatterer DOS given by
\begin{equation}
N_l^{(1)}=\frac{\sqrt{E_F}}{\pi}(2l+1)\int_0^{R_{MT}}R_l^2(r,E_F)r^2dr
\end{equation}
and the $\delta_l$ are the scattering phase shifts. Here $R_{MT}$ is the muffin-tin radius associated with atom type $j$ and $R_l$ is the scattering solution of the Schr\"{o}dinger equation. The phase shifts, which characterise the long-distance behaviour of the wavefunction, can be written in terms of the logarithmic derivative of the radial wavefunction
\begin{multline}
\label{phaseshifts}
\tan(\delta_l(R_{MT},E_F))=\\\frac{j'_l(kR_{MT})-j_l(kR_{MT})L_l(R_{MT},E_F)}{n'_l(kR_{MT})-n_l(kR_{MT})L_l(R_{MT},E_F)}
\end{multline}
where $k=\sqrt{E_F}$, $L_l=R'_l/R_l$ is the logarithmic derivative, $j_l$ are spherical Bessel functions and $n_l$ are Neumann functions. We can therefore directly calculate the logarithmic derivative and use Eq.\ \ref{phaseshifts} to obtain the phase shifts \cite{sakurai2017}.
\newline
\newline Since $M_j\braket{\omega_j^2}$ is often considerably smaller for hydrogen than for the other components, it is clear from Eq.\ \ref{totlambda} that the hydrogen atoms can provide a considerable fraction of $\lambda$ even if the Hopfield parameter of the other atom type is similar in magnitude. Calculating $\eta_H$ can therefore, in some cases, provide a cheap screening method for identifying potential high-$T_c$ hydrides. In particular, the average phonon frequencies for different structures are often similar when considering the same stoichiometry at the same pressure. If the average phonon frequencies are assumed to be exactly equivalent in such cases, we then arrive at a potential way of estimating $T_c$ ordering between structures, simply by considering $\eta_H$. It is in this context that we assess the utility of GG theory in this work.

\bibliography{references.bib}

\end{document}

\title{Predicting novel superconducting hydrides using machine learning approaches: supplementary information}

\author{Michael J. Hutcheon}
\email{mjh261@cam.ac.uk}
\affiliation
{
    Theory of Condensed Matter Group,
    Cavendish Laboratory,
    J.~J.~Thomson Avenue,
    Cambridge CB3 0HE,
    United Kingdom
}

\author{Alice M. Shipley}
\email{ams277@cam.ac.uk}
\affiliation
{
    Theory of Condensed Matter Group,
    Cavendish Laboratory,
    J.~J.~Thomson Avenue,
    Cambridge CB3 0HE,
    United Kingdom
}

\author{Richard J. Needs}
%\email{rn11@cam.ac.uk}
\affiliation
{
    Theory of Condensed Matter Group,
    Cavendish Laboratory,
    J.~J.~Thomson Avenue,
    Cambridge CB3 0HE,
    United Kingdom
}

\date{\today}

\maketitle

\section{Linear model}
To provide a baseline against which to test the neural network, we also carry out a linear regression on our binary hydride dataset, constructing the following models for critical temperature and operational pressure
\begin{align}
    T_c(d) &= \sum_{ij} c_{ij}^{(T)} d_i^{p_j}
    \\P(d) &= \sum_{ij} c_{ij}^{(P)} d_i^{p_j}
\end{align}
essentially building a linear combination of powers $p_j \in \{1, 2, 1/2, 1/4\}$ of the basic descriptors $d_i$ (introduced in the main text).

The goal of linear regression is to make a linear model of the relationship
\begin{equation}
    \underbrace{O_n}_{\text{Observable}} = O_n(\underbrace{f_{1,n}, f_{2,n}, \dots, f_{N,n}}_{\text{Features}})
\end{equation}
The linear model looks like
\begin{equation}
\label{eq:glr_model}
    \underbrace{P_n}_{\text{Prediction}} = \underbrace{\sum_i c_i f_{i,n}}_{\text{Linear model}} + \underbrace{\epsilon_n}_{Residual}
\end{equation}
where $c_i$ are the linear coefficients of the features. In matrix notation this looks like $P = fc + \epsilon$, where we call $f$ the \textit{feature matrix}. We pick the coefficient vector $c$ by minimizing the modulus of the residual vector; $c = \argmin_c |\epsilon|^2$. In order to simplify the model, we may bias the coefficient vector using Tikhonov regularization. This involves adding a cost function, $|Mc|^2$, which is large when the coefficient vector has many significant entries. The matrix $M$ is known as the Tikhonov matrix. This results in the minimization
\begin{equation}
\begin{aligned}
    c  &= \argmin_c \; |\epsilon|^2 + |Mc|^2 
     \\&= \argmin_c \; |fc - P|^2 + |Mc|^2 
     \\&\equiv \argmin_c \; \mathcal{L}(c)
\end{aligned}
\end{equation}
Using implied summation our objective function can be written as
\begin{equation}
    \mathcal{L}(c) = (f_{ij}c_j - P_i)(f_{ik}c_k - P_i) + M_{il}c_lM_{im}c_m
\end{equation}
Minimizing with respect to $c_n$ we require
\begin{equation}
    \frac{\partial \mathcal{L}}{\partial c_n} = 2[f_{in}(f_{ik}c_k - P_i) + M_{in}M_{ij}c_j] \overset{!}{=} 0
\end{equation}
In matrix notation this reads
\begin{equation}
    f^T(fc-p) + M^TMc = 0 \implies c = (f^Tf+M^TM)^{-1}f^Tp
\end{equation}
which gives us the optimal feature coefficients $c$ for the model in Eq.\ \ref{eq:glr_model}. 
In order to allow a direct comparison of the resulting coefficients, we normalize each column of the feature matrix to a mean of 1 before carrying out the regression. To reproduce the $L_2$ regularization that we apply to the (self-normalizing) neural network, we choose the (post-normalization) Tikhonov matrix as 10 times the identity matrix. The coefficients for the linear regression model discussed in the main text are given in Table \ref{tab:glr_model_hydrides}.

\begin{table*}
\centering
\subfigure[Pressure model]
{
\begin{tabular}{lrr}
    Term in expansion & Coefficient & Norm \\
    \hline
    Hydrogens/ion              &   16.91408 &    6.24775 \\
    Hydrogens/ion$^{0.5}$      &   16.43228 &    2.49955 \\
    Hydrogens/ion$^{0.25}$     &   15.24157 &    1.58100 \\
    S electrons$^{0.25}$       &   14.13379 &    1.16718 \\
    S electrons$^{0.5}$        &   13.85894 &    1.36230 \\
    S electrons                &   13.51666 &    1.85586 \\
    S electrons$^{2}$          &   13.43420 &    3.44420 \\
    Hydrogens/ion$^{2}$        &   12.43521 &   39.03435 \\
    VdW radius$^{0.25}$        &   12.23301 &    3.87678 \\
    VdW radius$^{0.5}$         &   11.49370 &   15.02940 \\
    VdW radius                 &   10.18405 &  225.88288 \\
    Ion Mass$^{2}$             &  -10.09502 & 15057.39153 \\
    P electrons                &    9.17555 &    0.39189 \\
    Ion atomic number$^{2}$    &   -8.16066 & 2560.72470 \\
    VdW radius$^{2}$           &    8.12073 & 51023.07678 \\
    F electrons                &   -7.88385 &    1.76577 \\
    Ion atomic number$^{0.25}$ &    7.45944 &    2.66714 \\
    Ion Mass$^{0.25}$          &    7.03580 &    3.32827 \\
    F electrons$^{0.5}$        &   -6.88904 &    1.32882 \\
    F electrons$^{0.25}$       &   -6.27614 &    1.15275 \\
    P electrons$^{0.5}$        &    4.73478 &    0.62601 \\
    D electrons$^{0.25}$       &    4.25008 &    1.26306 \\
    D electrons$^{0.5}$        &    3.91072 &    1.59532 \\
    Ion atomic number$^{0.5}$  &    3.72079 &    7.11362 \\
    D electrons                &    3.46107 &    2.54505 \\
    P electrons$^{0.25}$       &    3.33830 &    0.79121 \\
    Ion Mass$^{0.5}$           &    3.03423 &   11.07739 \\
    Ion Mass                   &   -2.21899 &  122.70856 \\
    F electrons$^{2}$          &    2.17818 &    3.11793 \\
    Ion atomic number          &   -1.16644 &   50.60360 \\
    D electrons$^{2}$          &   -1.06367 &    6.47725 \\
    P electrons$^{2}$          &   -0.34647 &    0.15358 \\
\end{tabular}
}
\hspace{2cm}
\subfigure[Temperature model]
{
\begin{tabular}{lrr}
    Term in expansion & Coefficient & Norm\\
    \hline
    Hydrogens/ion              &   16.34886 &    6.24775 \\
    Hydrogens/ion$^{2}$        &   15.72061 &   39.03435 \\
    F electrons                &  -14.62359 &    1.76577 \\
    F electrons$^{0.5}$        &  -12.34196 &    1.32882 \\
    Hydrogens/ion$^{0.5}$      &   11.42645 &    2.49955 \\
    F electrons$^{0.25}$       &  -11.15791 &    1.15275 \\
    S electrons$^{2}$          &    8.18184 &    3.44420 \\
    Hydrogens/ion$^{0.25}$     &    7.70346 &    1.58100 \\
    D electrons                &   -6.27340 &    2.54505 \\
    S electrons                &    5.97147 &    1.85586 \\
    D electrons$^{0.25}$       &    5.01341 &    1.26306 \\
    S electrons$^{0.5}$        &    4.45893 &    1.36230 \\
    VdW radius$^{2}$           &    4.12143 & 51023.07678 \\
    VdW radius                 &    3.58428 &  225.88288 \\
    S electrons$^{0.25}$       &    3.55632 &    1.16718 \\
    Ion Mass$^{0.25}$          &    3.33243 &    3.32827 \\
    Ion atomic number$^{0.25}$ &    3.32475 &    2.66714 \\
    Ion atomic number$^{0.5}$  &    3.28585 &    7.11362 \\
    VdW radius$^{0.5}$         &    3.26998 &   15.02940 \\
    Ion Mass$^{0.5}$           &    3.25892 &   11.07739 \\
    Ion Mass$^{2}$             &   -3.24468 & 15057.39153 \\
    VdW radius$^{0.25}$        &    3.10714 &    3.87678 \\
    P electrons                &    2.72494 &    0.39189 \\
    Ion atomic number          &    2.35759 &   50.60360 \\
    Ion atomic number$^{2}$    &   -2.20360 & 2560.72470 \\
    Ion Mass                   &    2.15283 &  122.70856 \\
    F electrons$^{2}$          &    1.90288 &    3.11793 \\
    D electrons$^{2}$          &   -0.87843 &    6.47725 \\
    P electrons$^{0.25}$       &   -0.39708 &    0.79121 \\
    D electrons$^{0.5}$        &    0.25192 &    1.59532 \\
    P electrons$^{0.5}$        &    0.13546 &    0.62601 \\
    P electrons$^{2}$          &   -0.00791 &    0.15358 \\
\end{tabular}
}
\caption{Linear regression model for the critical temperature and corresponding operational pressure of hydrides. Each parameter is normalized prior to the regression by dividing by the average value for the dataset (right-hand column).}
\label{tab:glr_model_hydrides}
\end{table*}

\section{Convex hulls}
\noindent We constructed (static-lattice) convex hulls using AIRSS \cite{pickard2011} and qhull \cite{barber1996} for various hydride systems (at 50, 100 and 200\ GPa). As detailed in the main text, we produced convex hulls for systems that were indicated as promising for superconductivity closer to ambient conditions by our machine learning model and linear regression. The systems we focus on here are Cs-H, Ra-H and Rb-H and the hulls produced inform our further structure searches.

\begin{figure}[H]
    \centering
    \includegraphics[width=0.45\textwidth]{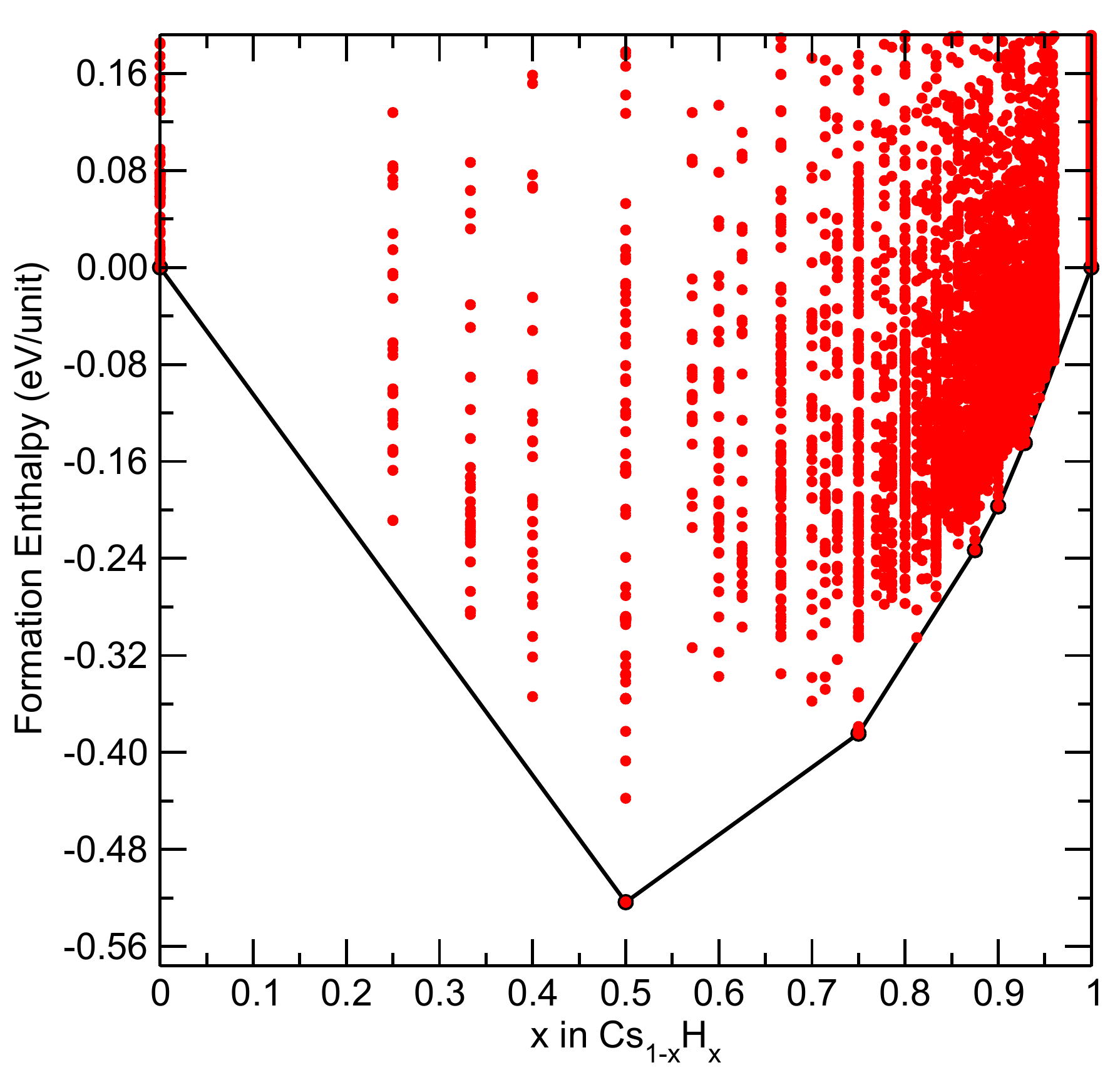}
    \caption{Convex hull for Cs-H system at 50\ GPa}
    \label{fig:csh_hull50}
\end{figure}

As shown in Fig.\ \ref{fig:csh_hull50}, CsH, CsH$_3$, CsH$_7$, CsH$_9$ and CsH$_{13}$ are on the hull at 50\ GPa. There are many compositions that are very close to the hull at this pressure, including Cs$_3$H$_{13}$, CsH$_{24}$, CsH$_{17}$, CsH$_{14}$, CsH$_{12}$, and CsH$_{15}$.

\begin{figure}[H]
    \centering
    \includegraphics[width=0.45\textwidth]{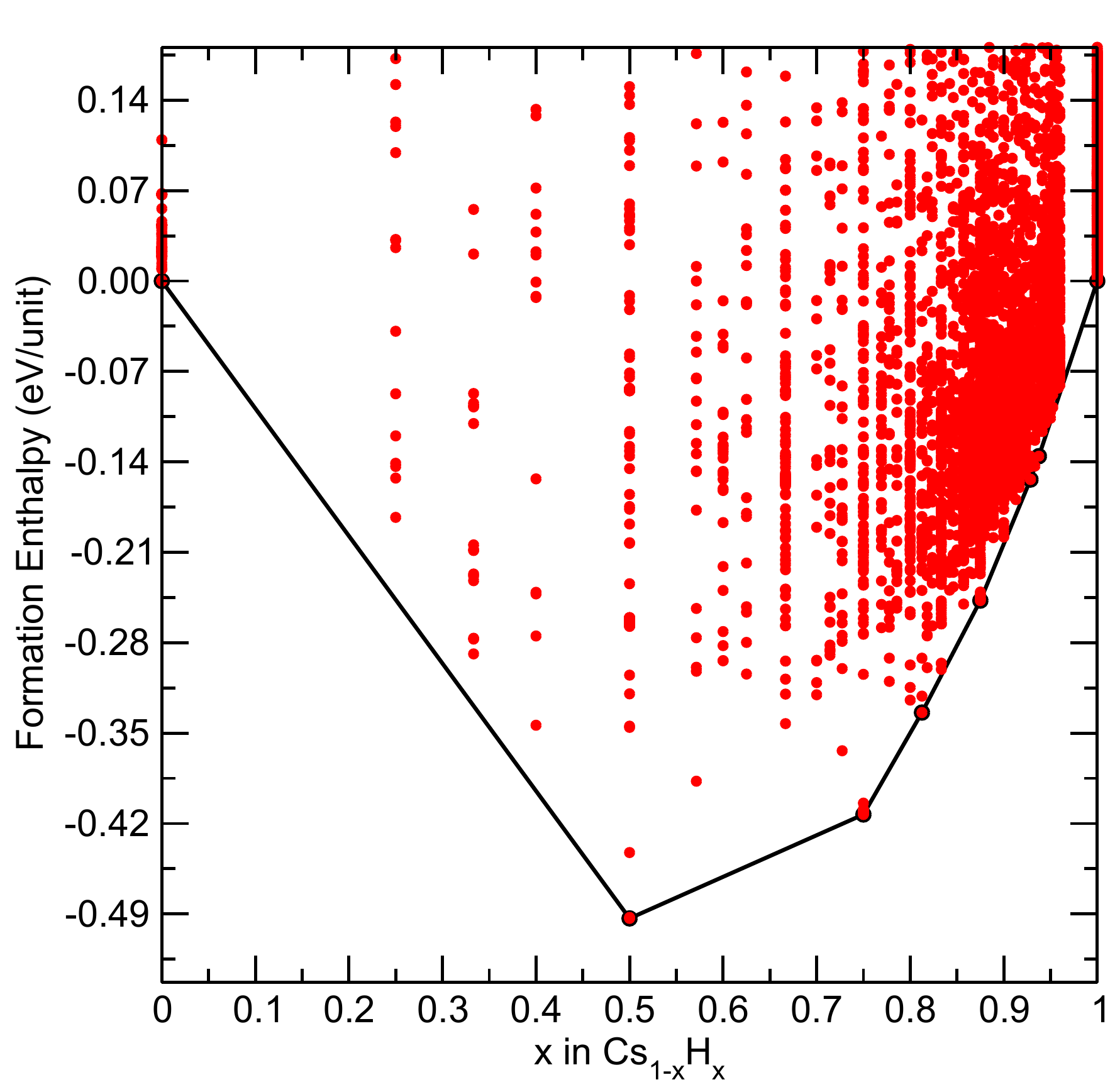}
    \caption{Convex hull for Cs-H system at 100\ GPa}
    \label{fig:csh_hull100}
\end{figure}

\begin{figure}[H]
    \centering
    \includegraphics[width=0.45\textwidth]{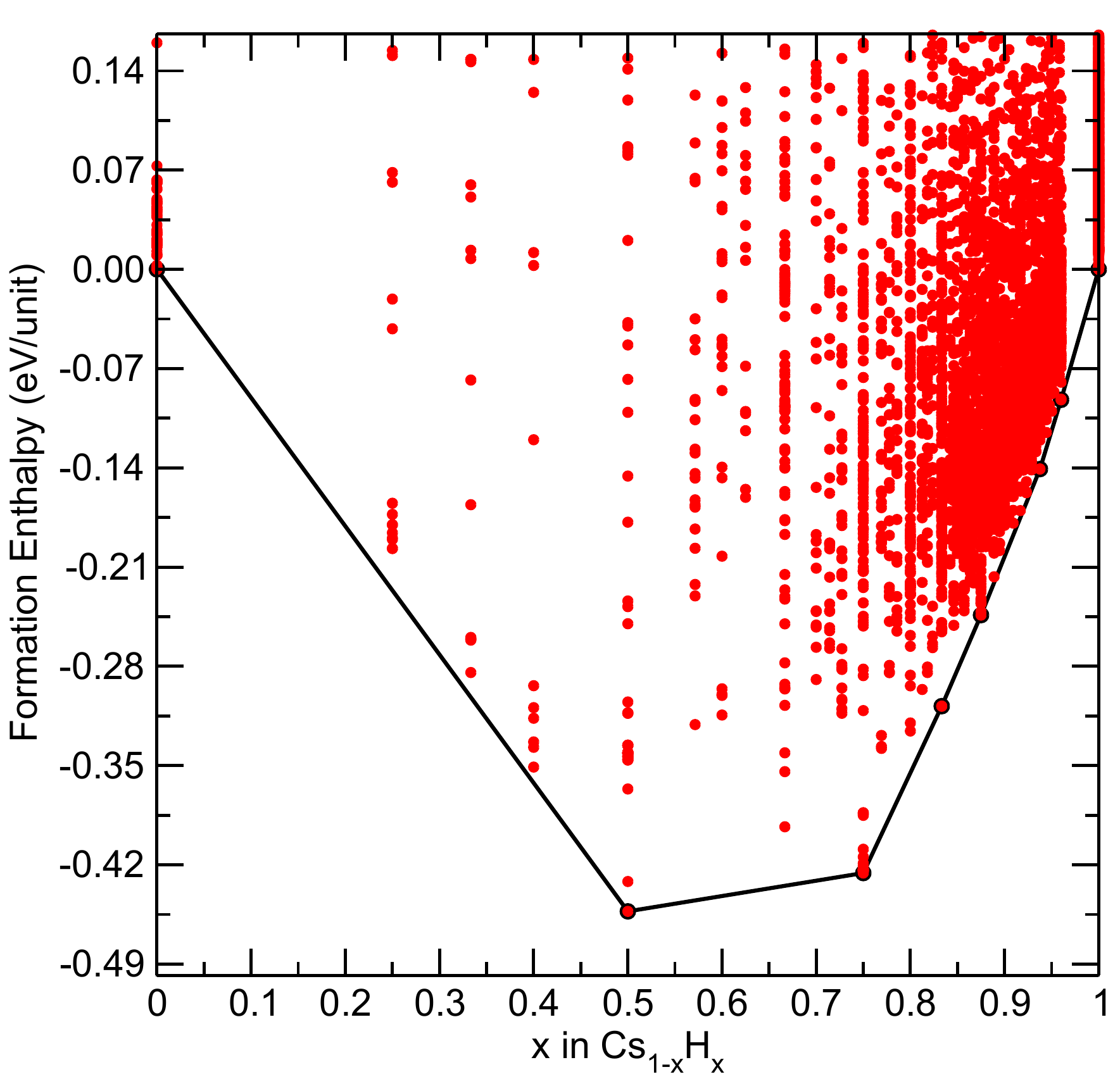}
    \caption{Convex hull for Cs-H system at 200\ GPa}
    \label{fig:csh_hull200}
\end{figure}

As shown in Fig.\ \ref{fig:csh_hull100}, CsH, CsH$_3$, Cs$_3$H$_{13}$, CsH$_7$, CsH$_{13}$ and CsH$_{15}$ are on the hull at 100\ GPa. We can see from Fig.\ \ref{fig:csh_hull200} that CsH$_7$ and CsH$_{15}$ remain on the hull at 200\ GPa, while CsH$_{13}$ is found slightly above it. We also note that CsH$_5$ is on the hull at 200\ GPa. We chose to investigate CsH$_5$, CsH$_7$, CsH$_{13}$ and CsH$_{15}$ further.

\begin{figure}[H]
    \centering
    \includegraphics[width=0.45\textwidth]{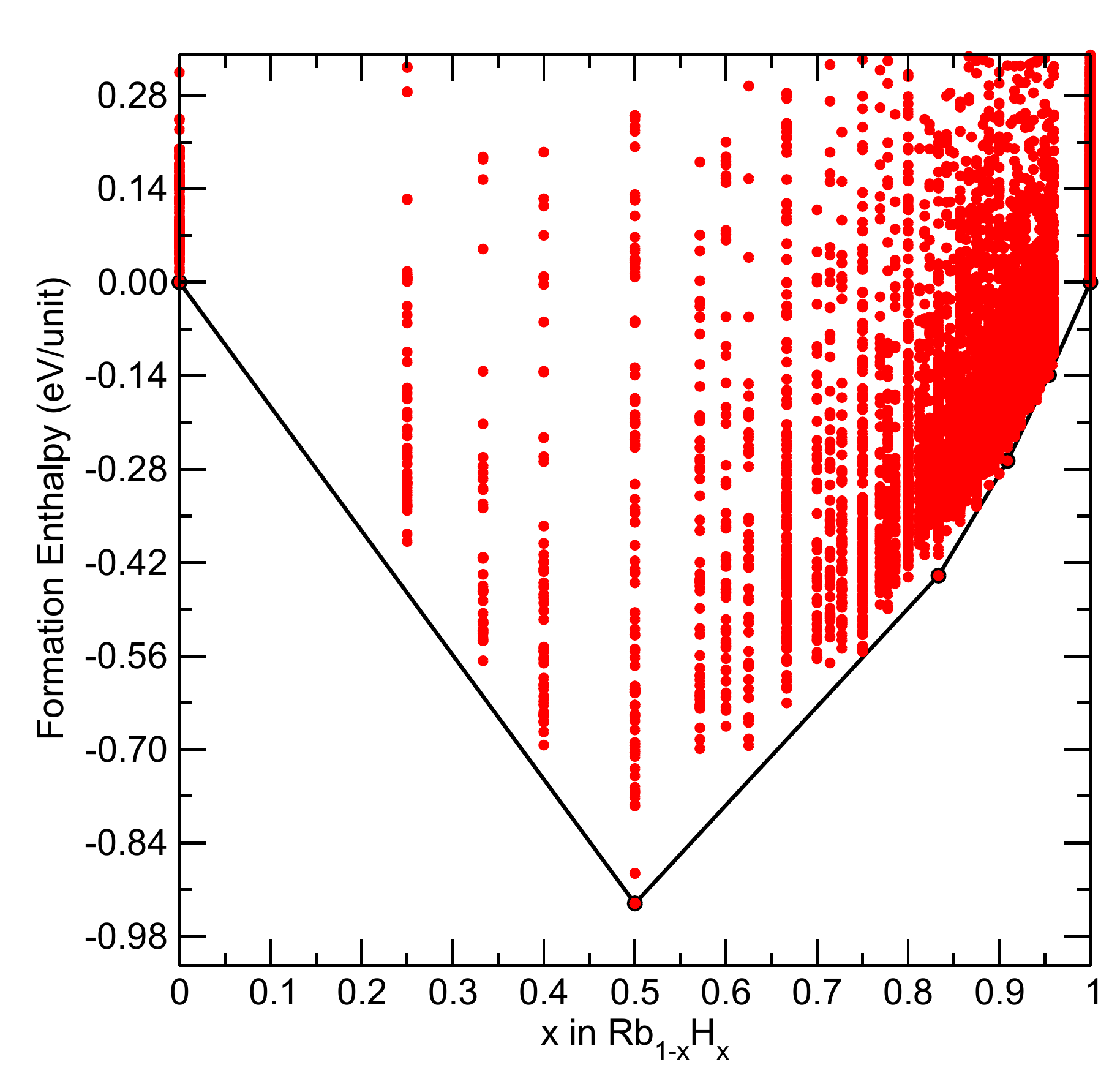}
    \caption{Convex hull for Rb-H system at 50\ GPa}
    \label{fig:rbh_hull50}
\end{figure}

As shown in Fig.\ \ref{fig:rbh_hull50}, RbH, RbH$_5$, RbH$_{10}$ and RbH$_{21}$ are on the hull at 50\ GPa. There are many compositions close to the hull at this pressure, including RbH$_{16}$, RbH$_9$, RbH$_{23}$, RbH$_{19}$, RbH$_{17}$, RbH$_8$, RbH$_7$, RbH$_3$, RbH$_{14}$ and RbH$_{12}$.

\begin{figure}[H]
    \centering
    \includegraphics[width=0.45\textwidth]{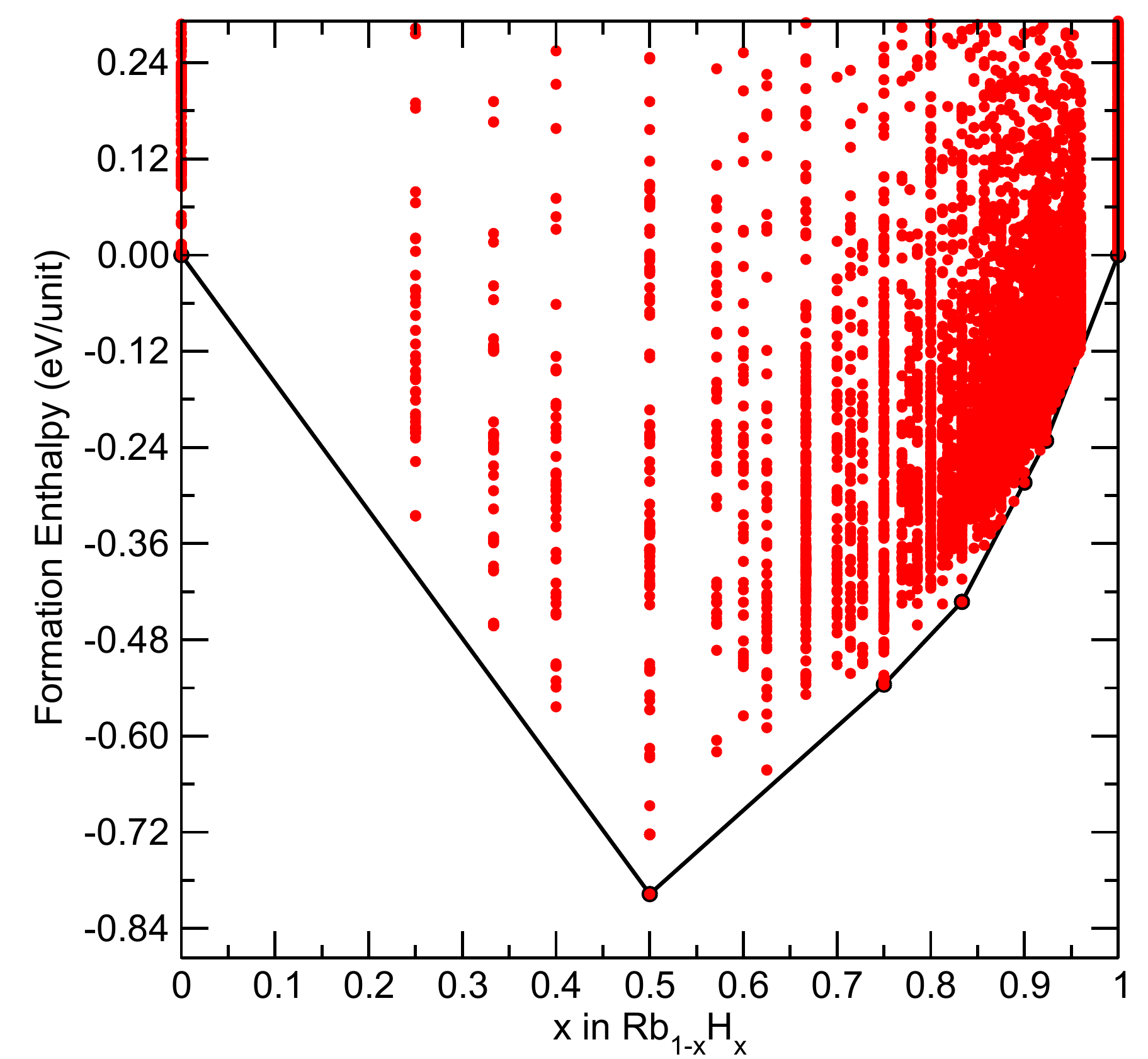}
    \caption{Convex hull for Rb-H system at 100\ GPa}
    \label{fig:rbh_hull100}
\end{figure}

\begin{figure}[H]
    \centering
    \includegraphics[width=0.45\textwidth]{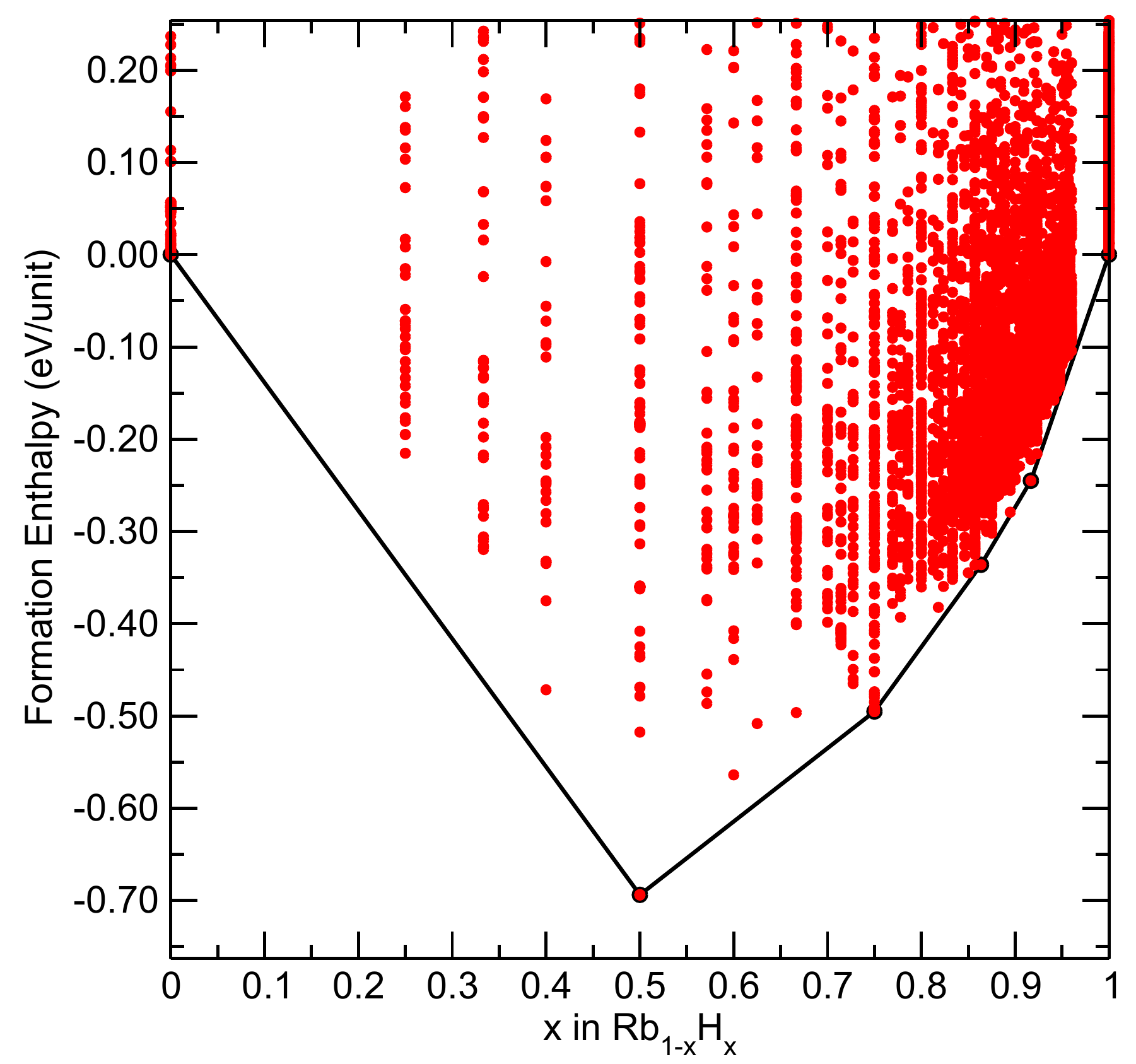}
    \caption{Convex hull for Rb-H system at 200\ GPa}
    \label{fig:rbh_hull200}
\end{figure}

As shown in Fig.\ \ref{fig:rbh_hull100}, RbH, RbH$_3$, RbH$_5$, RbH$_9$ and RbH$_{12}$ are on the hull at 100\ GPa. At 200\ GPa (Fig.\ \ref{fig:rbh_hull200}), RbH$_{12}$ is above but close to the hull, while RbH$_5$ and RbH$_9$ are found further away. RbH$_3$ and RbH$_{11}$ are also on the hull at 200\ GPa. We chose to investigate RbH$_3$, RbH$_5$, RbH$_9$, RbH$_{11}$ and RbH$_{12}$ further.

\begin{figure}[H]
    \centering
    \includegraphics[width=0.45\textwidth]{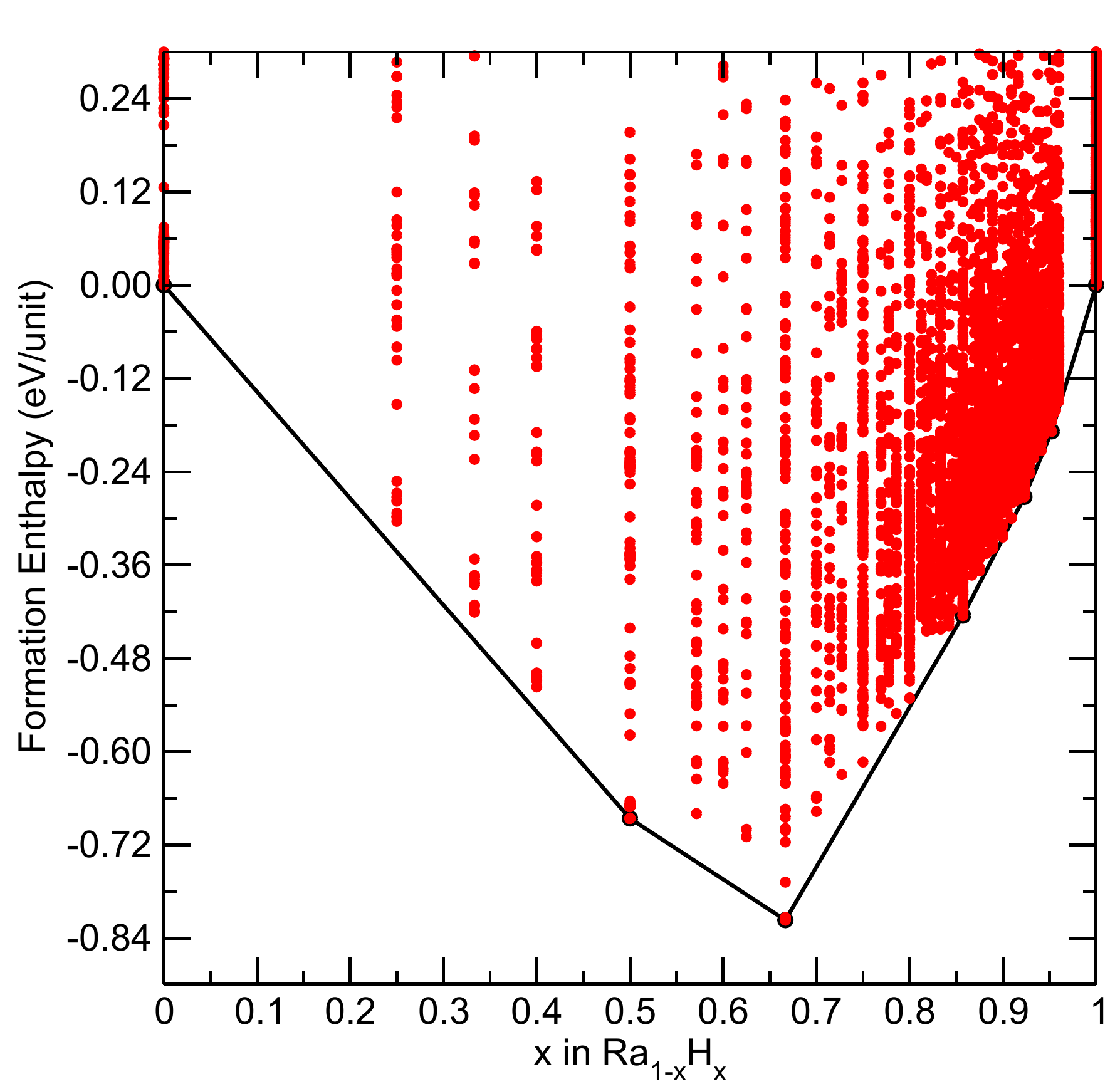}
    \caption{Convex hull for Ra-H system at 100\ GPa}
    \label{fig:rah_hull100}
\end{figure}
Radium hydrides were also indicated as promising candidates by the models. However, these hydrides were eliminated from further calculations as all isotopes of radium are radioactive. Although radium hydrides were not investigated any further in this work, the Ra-H convex hull at 100\ GPa is shown in Fig.\ \ref{fig:rah_hull100}.

\section{Enthalpy plots and electronic density of states}

After constructing convex hulls and deciding which stoichiometries to investigate, we performed further AIRSS searches. From these searches we then selected the lowest enthalpy candidates and relaxed these structures within \textsc{quantum espresso} \cite{QE-2009,QE-2017} using more stringent parameters (as detailed in the main text) at a range of pressures. These calculations (and the following electron-phonon calculations) all used scalar-relativistic, ultrasoft PBE pseudopotentials downloaded from \\ \url{https://www.quantum-espresso.org/pseudopotentials/ps-library}. Before performing expensive electron-phonon calculations to assess potential superconductivity, we also wanted to ensure that the materials were metallic at the pressures of interest. In the following Tables we therefore present the electronic density of states (DOS) at the Fermi level ($E_F$) for each of the structures at 50\ GPa and 150\ GPa, calculated using \textsc{quantum espresso}.

\begin{figure}
    \centering
    \includegraphics[width=\columnwidth]{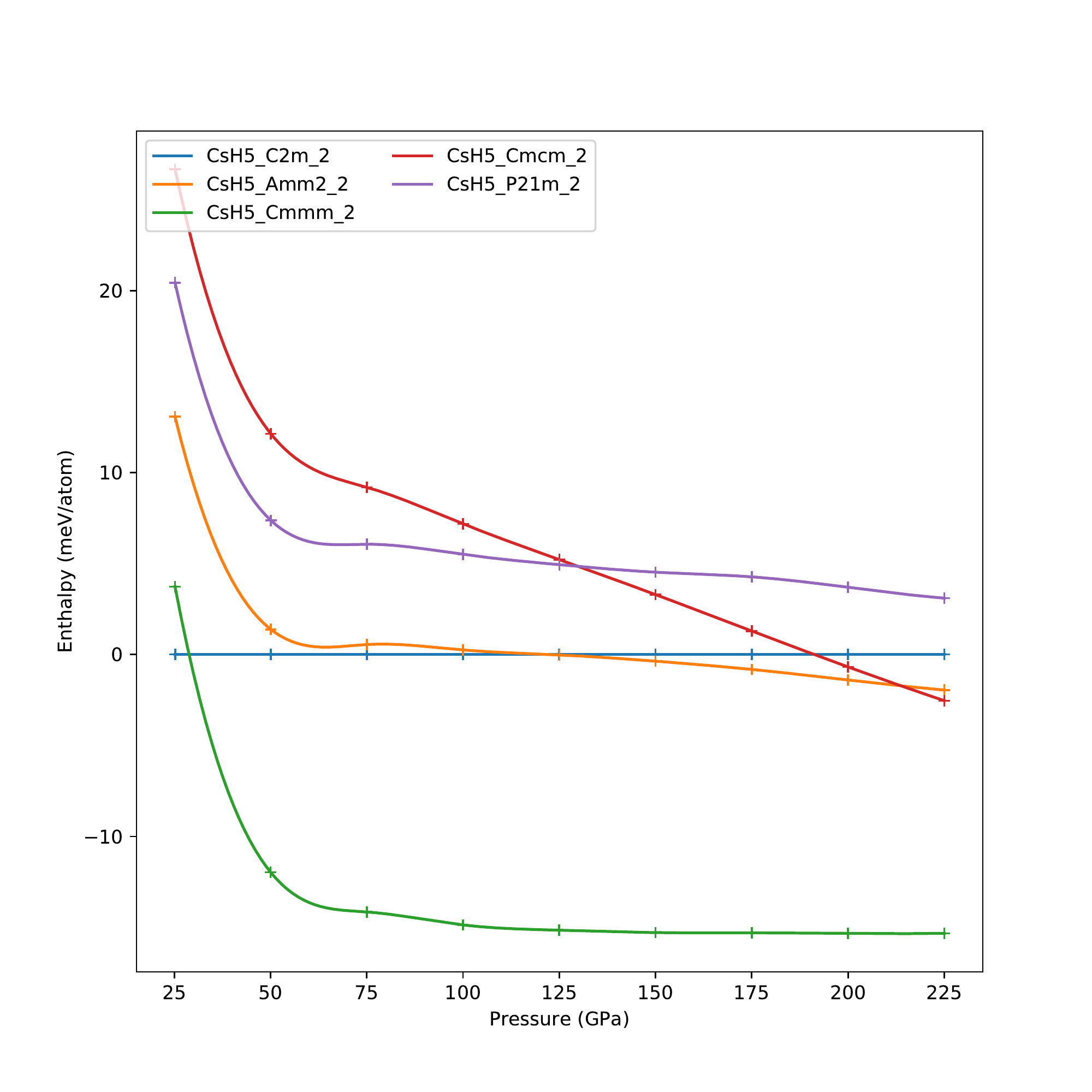}
    \caption{Enthalpy vs pressure plot for CsH$_5$ structures from AIRSS}
    \label{fig:csh5_enth}
\end{figure}

\begin{figure}
    \centering
    \includegraphics[width=\columnwidth]{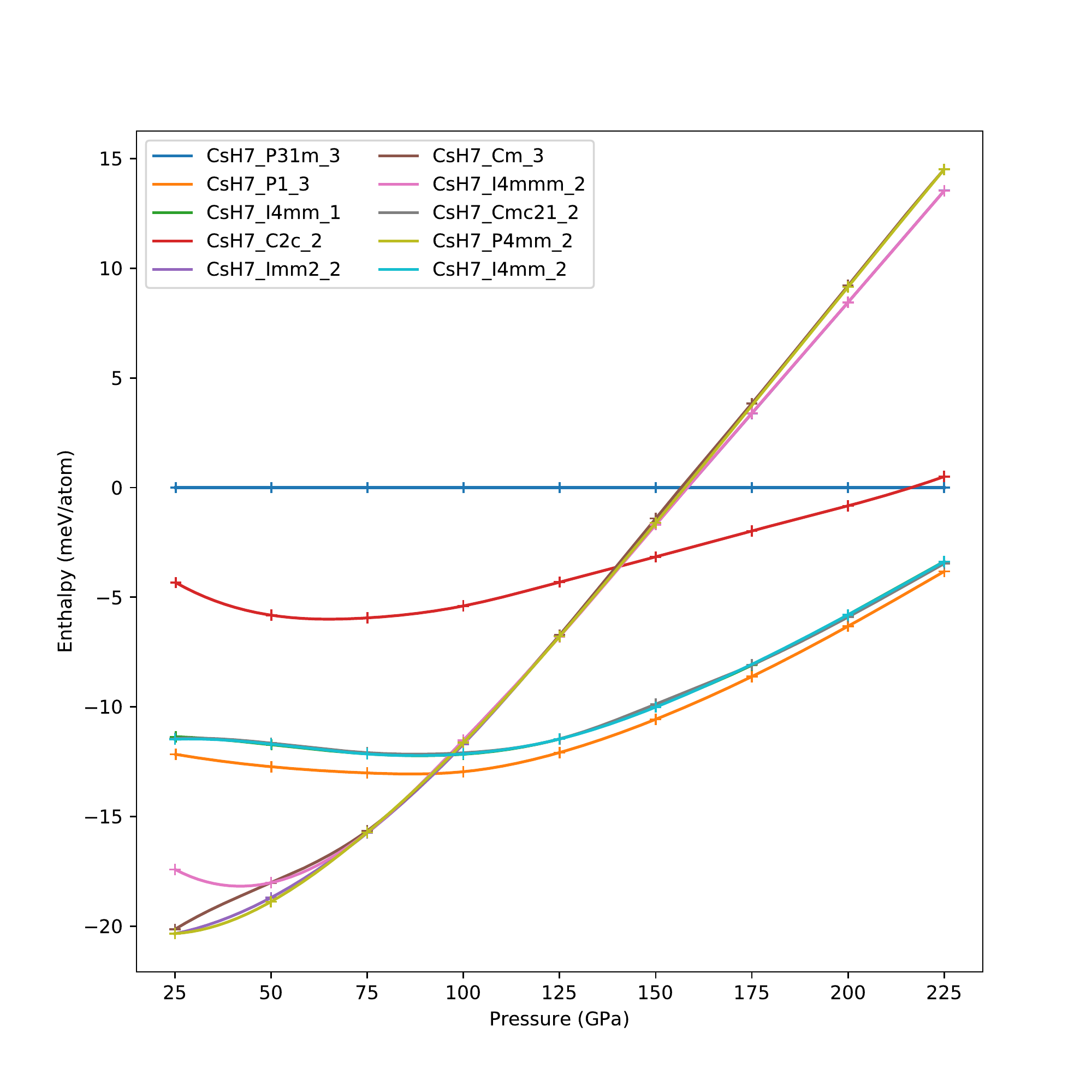}
    \caption{Enthalpy vs pressure plot for CsH$_7$ structures from AIRSS}
    \label{fig:csh7_enth}
\end{figure}

\begin{figure}
    \centering
    \includegraphics[width=\columnwidth]{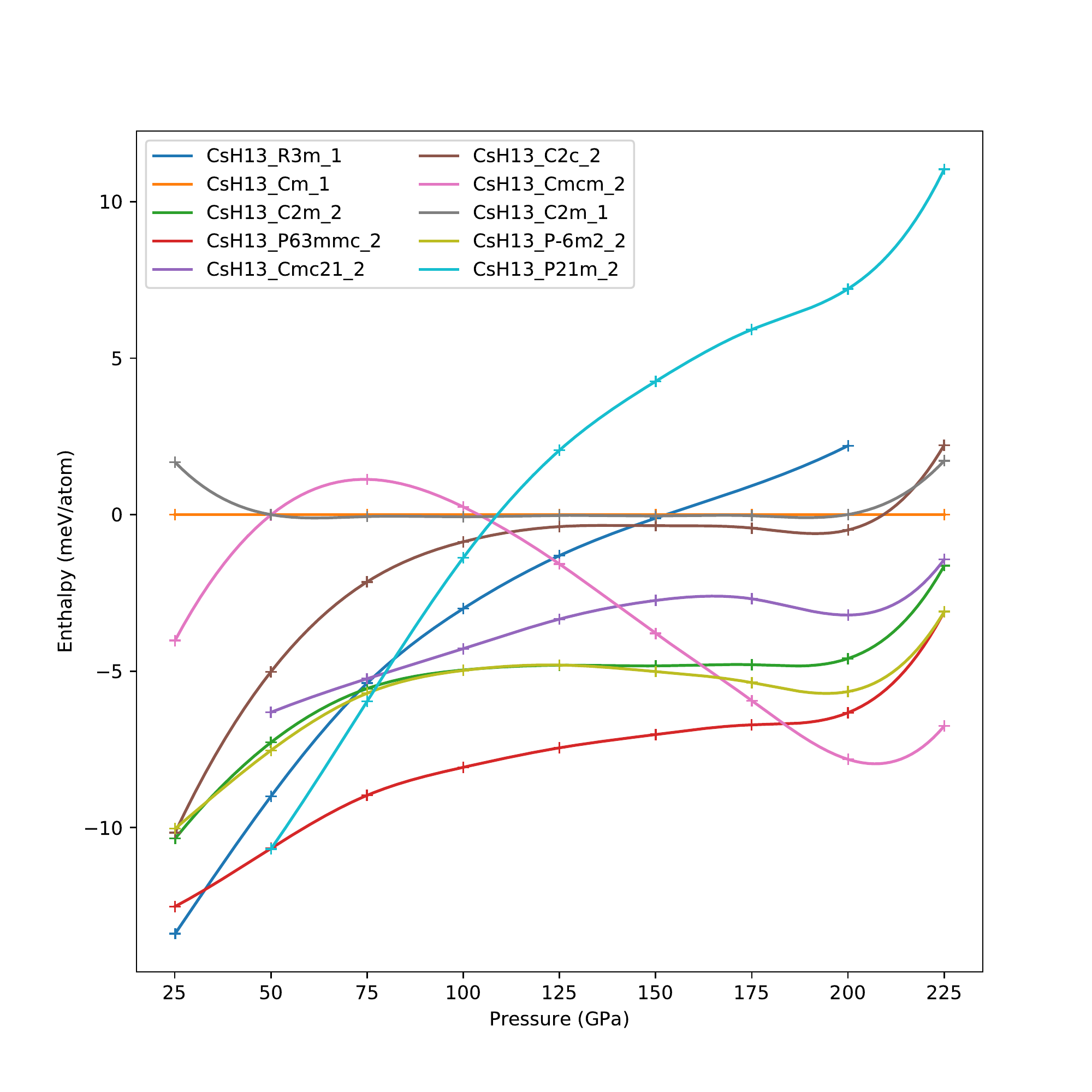}
    \caption{Enthalpy vs pressure plot for CsH$_{13}$ structures from AIRSS}
    \label{fig:csh13_enth}
\end{figure}

\begin{figure}
    \centering
    \includegraphics[width=\columnwidth]{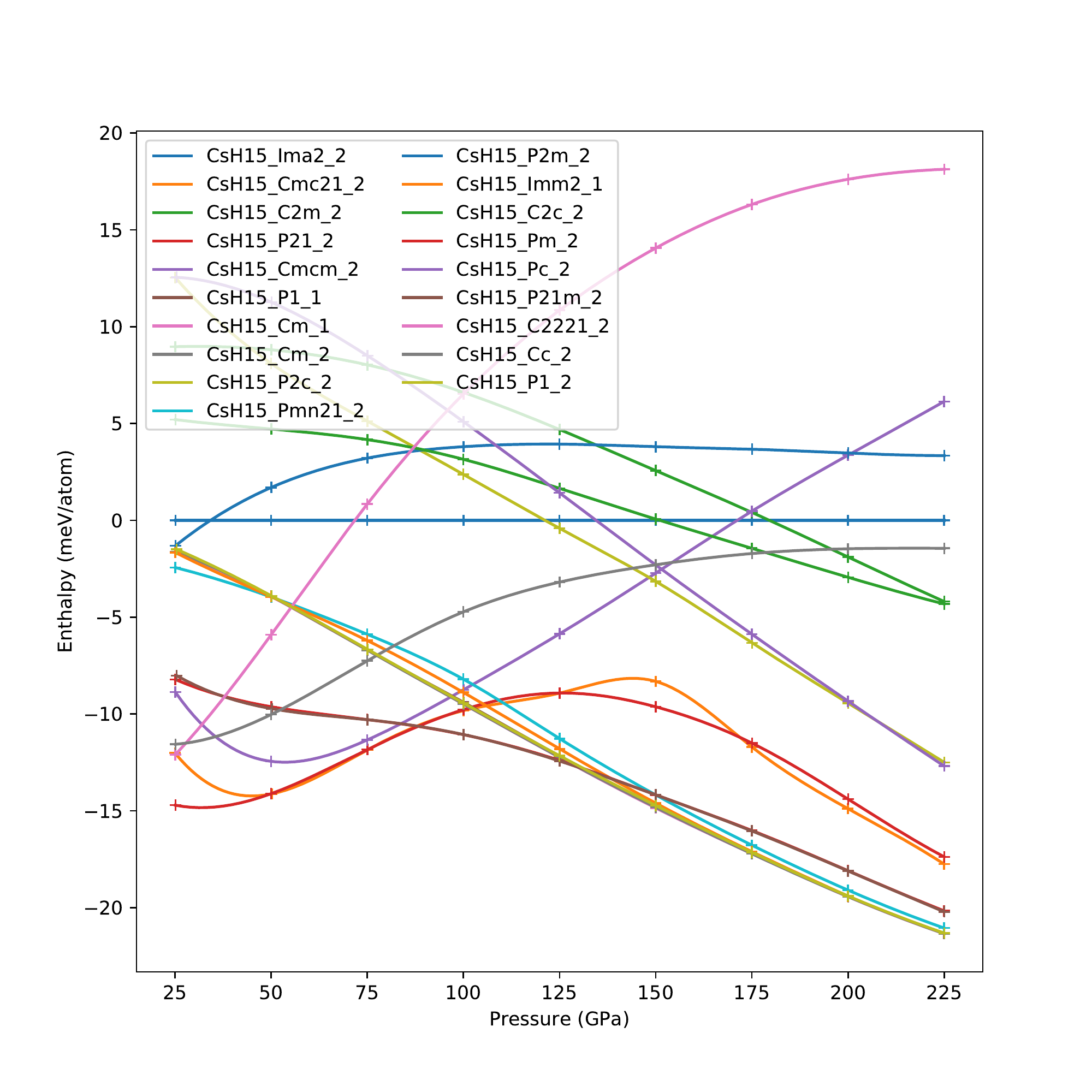}
    \caption{Enthalpy vs pressure plot for CsH$_{15}$ structures from AIRSS}
    \label{fig:csh15_enth}
\end{figure}

\begin{figure}
    \centering
    \includegraphics[width=\columnwidth]{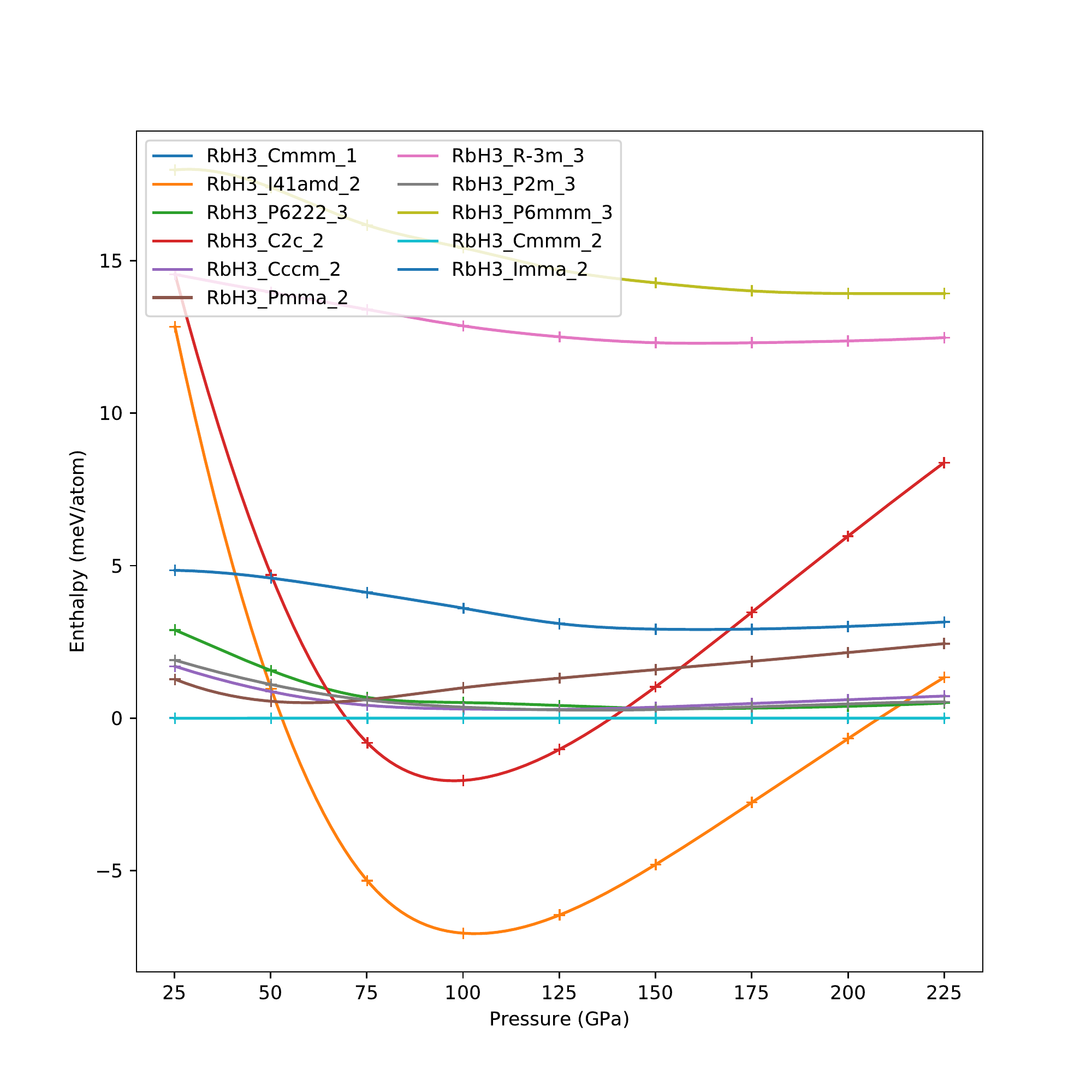}
    \caption{Enthalpy vs pressure plot for RbH$_3$ structures from AIRSS}
    \label{fig:rbh3_enth}
\end{figure}

\begin{figure}
    \centering
    \includegraphics[width=\columnwidth]{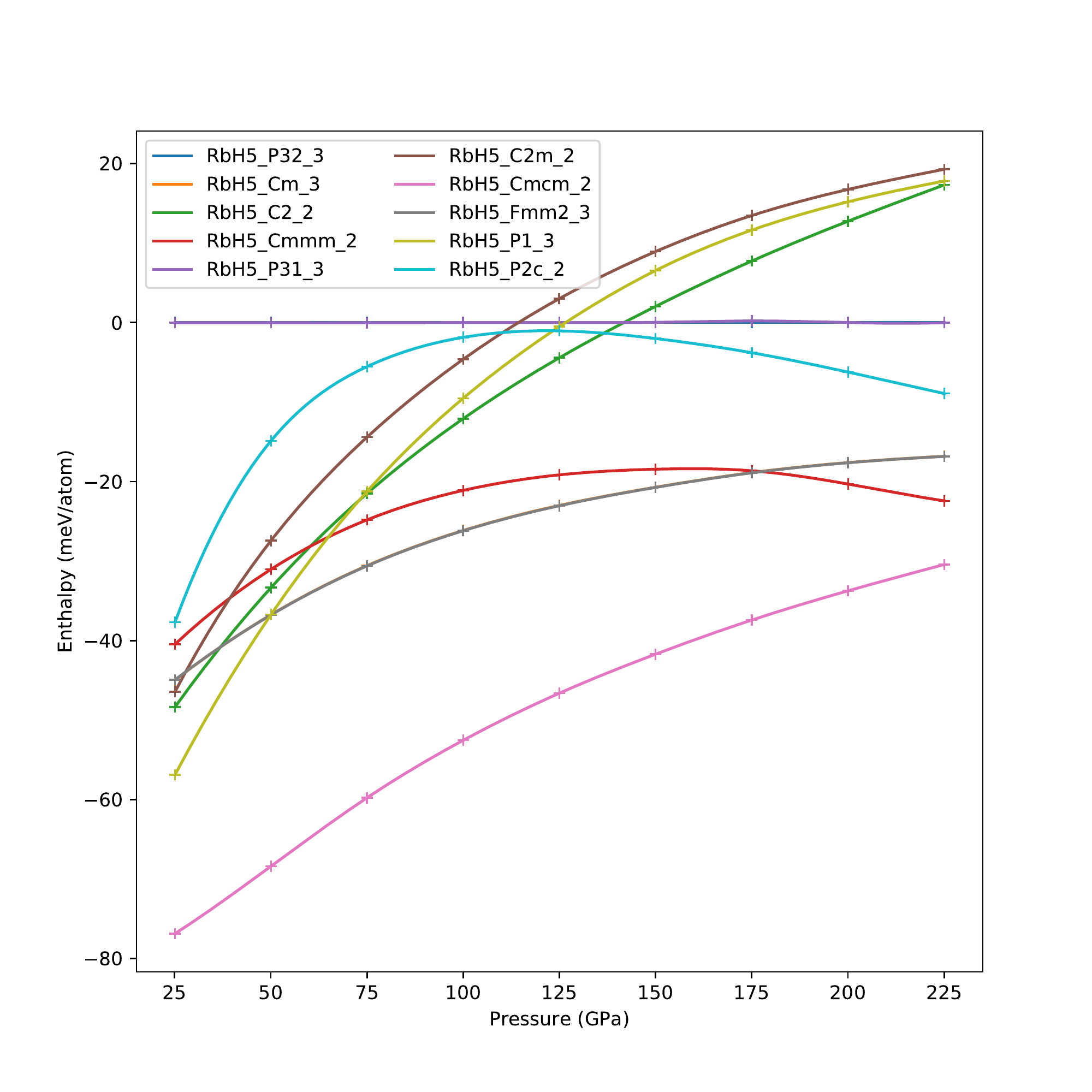}
    \caption{Enthalpy vs pressure plot for RbH$_5$ structures from AIRSS}
    \label{fig:rbh5_enth}
\end{figure}

\begin{figure}
    \centering
    \includegraphics[width=\columnwidth]{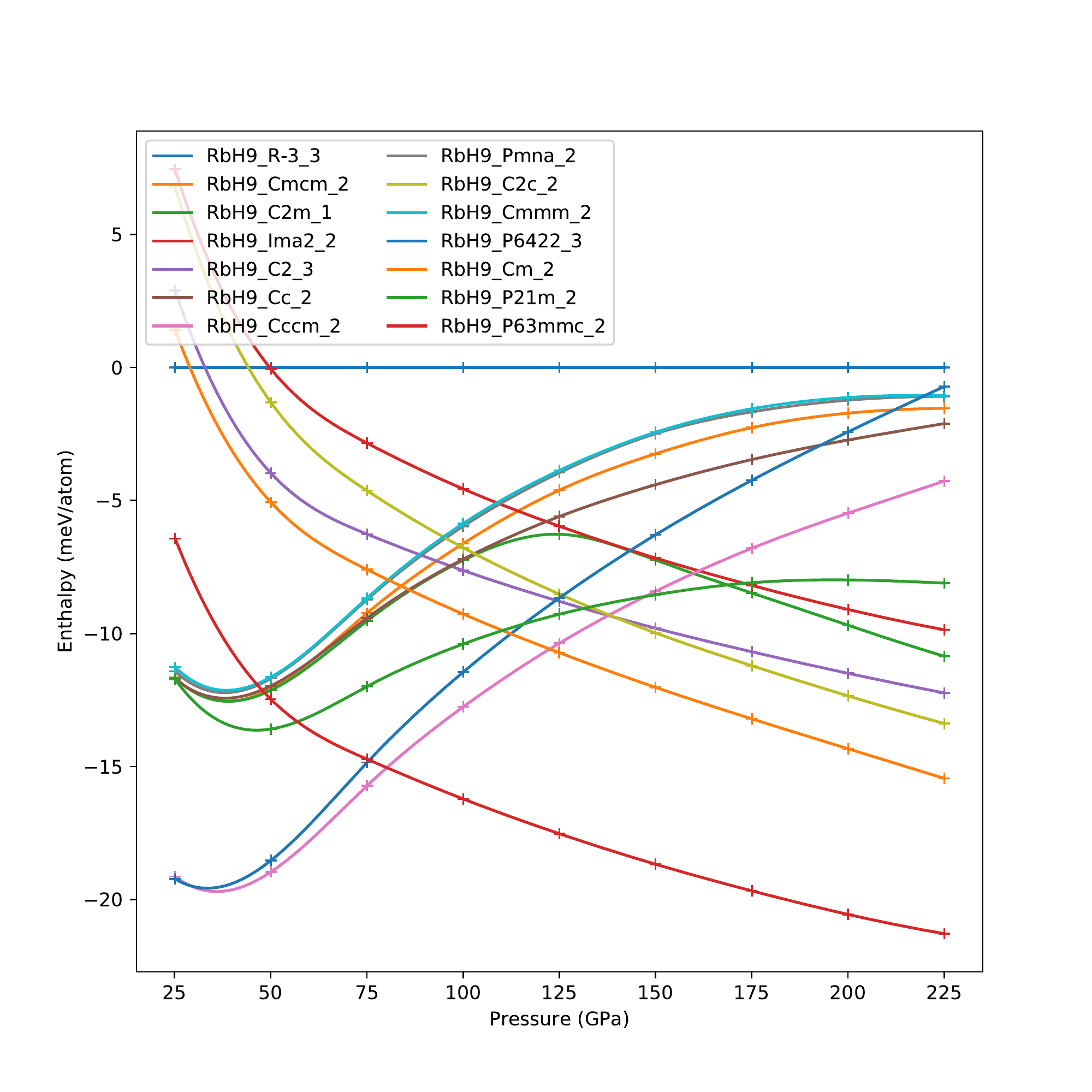}
    \caption{Enthalpy vs pressure plot for RbH$_9$ structures from AIRSS}
    \label{fig:rbh9_enth}
\end{figure}

\begin{figure}
    \centering
    \includegraphics[width=\columnwidth]{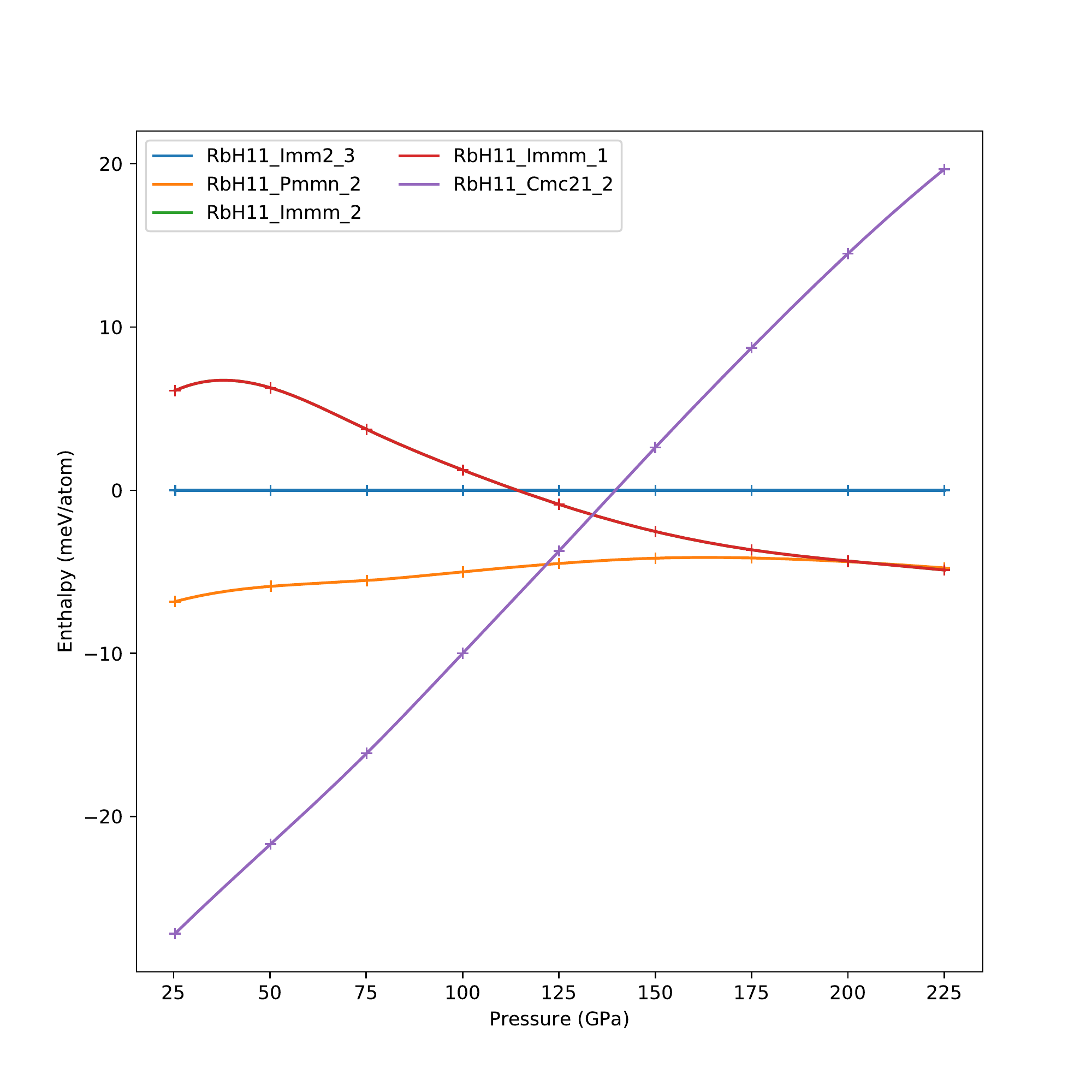}
    \caption{Enthalpy vs pressure plot for RbH$_{11}$ structures from AIRSS}
    \label{fig:rbh11_enth}
\end{figure}

\begin{figure}
    \centering
    \includegraphics[width=\columnwidth]{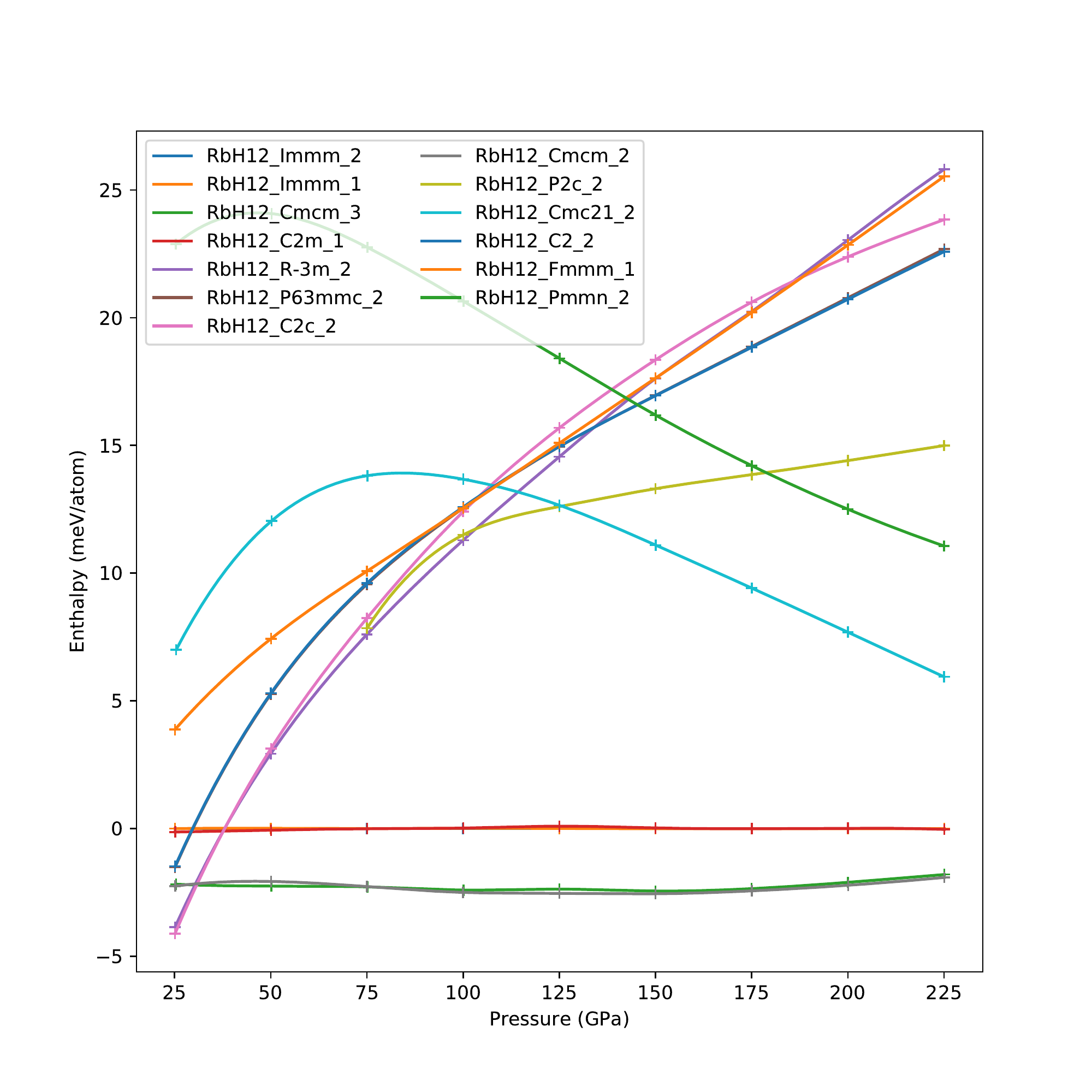}
    \caption{Enthalpy vs pressure plot for RbH$_{12}$ structures from AIRSS}
    \label{fig:rbh12_enth}
\end{figure}

\clearpage
\begin{widetext}
\begin{center}
\begin{tabular}{ |p{2cm}|p{2.5cm}|p{5.5cm}|p{5.75cm}|  }
 \hline
 \multicolumn{4}{|c|}{CsH$_5$ structures} \\
 \hline
 Symmetry& Formula units & DOS($E_F$)(states/eV/f.u.) at 50 GPa & DOS($E_F$)(states/eV/f.u.) at 150 GPa\\
 \hline
 $Amm2$ &   2   &   0.5977E-03  &0.2622\\
 $C2/m$ &   2   &   0.1092E-07  &0.1559E-03\\
 $Cmcm$ &   2   &   0.8085E-10  &0.2435\\
 $Cmmm$ &   2   &   0.4350E-04  &0.1667E-01\\
 $P2_1/m$   &   2   &   0.6636E-09  &0.1352\\
 \hline
\end{tabular}
\end{center}

\begin{center}
\begin{tabular}{ |p{2cm}|p{2.5cm}|p{5.5cm}|p{5.75cm}|  }
 \hline
 \multicolumn{4}{|c|}{CsH$_7$ structures} \\
 \hline
 Symmetry& Formula units & DOS($E_F$)(states/eV/f.u.) at 50GPa & DOS($E_F$)(states/eV/f.u.) at 150GPa\\
 \hline
 $Cm$   &   3   &   0.100005    &0.41645\\
 $Cmc2_1$   &   2  &    0.1763  &0.3571\\
 $I4mm$ &   2   &   0.7785E-01  &0.1357\\
 $I4mmm$    &   2   &   0.6736E-01  &0.2685\\
 $Imm2$ &   2   &   0.8825E-02  &0.29315\\
 $P4mm$ &   2   &   0.016045    &0.1736\\
 $C2/c$ &   2   &   0.3339E-07  &0.3146\\
 $I4mm$ &   1   &   0.7988E-01  &0.1812\\
 $P1$   &   3   &   0.1898  &0.6102\\
 $P3_1/m$   &   3   &   0.6961E-07  &0.7867E-02\\
 \hline
\end{tabular}
\end{center}

\begin{center}
\begin{tabular}{ |p{2cm}|p{2.5cm}|p{5.5cm}|p{5.75cm}|  }
 \hline
 \multicolumn{4}{|c|}{CsH$_{13}$ structures} \\
 \hline
 Symmetry& Formula units & DOS($E_F$)(states/eV/f.u.) at 50GPa & DOS($E_F$)(states/eV/f.u.) at 150GPa\\
 \hline
 $C2/c$   & 2    & 3.7610312E-28 & 0.00013395\\
 $C2/m$&   1  & 1.13605E-06 & 0.1712\\
 $C2/m$ &2 &   0.3994E-21 & 0.00002664\\
 $Cm$    &1 &   0.5642E-06 & 0.1891\\
 $Cmcm$&   2  & 2.1535181E-27 & 0.030615\\
 $P\bar{6}m2$& 2  & 0.7362E-05 & 0.1161\\
 $P2_1/m$& 2  & 0.7736E-04 & 0.5526\\
 $P6_3/mmc$    &2 &   0.7946E-15 & 0.1680\\
 $R3m$&   1  & 0.49935E-03 & 0.4264\\
 $Cmc2_1$&   2  & -- & 0.6071\\
 \hline
\end{tabular}
\end{center}

\begin{center}
\begin{tabular}{ |p{2cm}|p{2.5cm}|p{5.5cm}|p{5.75cm}|  }
 \hline
 \multicolumn{4}{|c|}{CsH$_{15}$ structures} \\
 \hline
 Symmetry& Formula units & DOS($E_F$)(states/eV/f.u.) at 50GPa & DOS($E_F$)(states/eV/f.u.) at 150GPa\\
 \hline
 $C2/m$	&	2	&	0.4644E-07 &0.031865\\
 $Cm$	&	1	&	2.9626484E-21 &0.3559E-02\\
 $Cm$	&	2	&	2.5831273E-21 &0.6947E-02\\
 $Cmcm$	&	2	&	0.9725E-10 &0.02838\\
 $Ima2$	&	2	&	0.1350E-19 &0.2176\\
 $Imm2$	&	1	&	7.3203181E-22 &0.136455E-04\\
 $P1$	&	1	&	0.7504E-26 &0.7516E-02\\
 $P2_1$	&	2	&	0.3003E-04 &0.7784\\
 $P2/c$	&	2	&	0.7705E-08 &0.1093E-05\\
 $C2221$ &  2   &   0.1044E-15 &0.0058975\\
 $C2/c$ &   2   &   0.5056E-06 &0.2833E-02\\
 $Cc$  &    2   &   0.2747E-26 &0.0029778\\
 $P2/m$ &   2   &   0.1263E-21 &0.10651\\
 $Pm$   &   2   &   0.109812E-05 &0.4108\\
 $Pmn2_1$   &   2   &   0.8539E-53 &5.1819705E-14\\
 $Cmc2_1$ & 2   &   0.3265E-04  &0.9962\\
 $P1$   &   2   &   0.1520E-30  &0.2716E-18\\
 $P2_1/m$   &   2   &   0.7671E-04  &0.7977E-14\\
 $Pc$   &   2   &   0.1534E-16  &0.5661E-07\\
 \hline
\end{tabular}
\end{center}

\begin{center}
\begin{tabular}{ |p{2cm}|p{2.5cm}|p{5.5cm}|p{5.75cm}|  }
 \hline
 \multicolumn{4}{|c|}{RbH$_3$ structures} \\
 \hline
 Symmetry& Formula units & DOS($E_F$)(states/eV/f.u.) at 50GPa & DOS($E_F$)(states/eV/f.u.) at 150GPa\\
 \hline
 $C2/c$ &   2   &   0.1127E-08  &0.3533E-41\\
 $Cccm$ &   2   &   0.8055E-01  &0.2819\\
 $Cmmm$ &   1   &   0.3865E-01  &0.1284\\
 $I4_1/amd$ &   2   &   0.4432E-01  &0.3181E-03\\
 $Imma$ &   2   &   0.9423E-01  &0.2798\\
 $P2/m$ &   3   &   0.1244  &0.4055\\
 $P6222$    &   3   &   0.1189  &0.3862\\
 $P6mmm$    &   3   &   0.2180  &0.5852\\
 $Pmma$ &   2   &   0.6661E-01  &0.2889\\
 $R\bar{3}m$    &   3   &   0.1720  &--\\
 \hline
\end{tabular}
\end{center}

\begin{center}
\begin{tabular}{ |p{2cm}|p{2.5cm}|p{5.5cm}|p{5.75cm}|  }
 \hline
 \multicolumn{4}{|c|}{RbH$_5$ structures} \\
 \hline
 Symmetry& Formula units & DOS($E_F$)(states/eV/f.u.) at 50GPa & DOS($E_F$)(states/eV/f.u.) at 150GPa\\
 \hline
 $C2$   &   2   &   2.1730038E-35  &0.1004E-13\\
 $C2/m$ &   2   &   0.1588E-25  &0.1780\\
 $Cm$   &   3   &   2.926505E-35  &0.2993E-04\\
 $Cmcm$ &   2   &   9.9900005E-57  &1.347721E-16\\
 $Cmmm$ &   2   &   0.1123E-14  &0.1138\\
 $Fmm2$ &   3   &   9.8400152E-36  &--\\
 $P1$   &   3   &   0.6238E-01  &0.2623\\
 $P2/c$ &   2   &   0.216509E-04  &0.1228\\
 $P31$  &   3   &   0.3055E-19  & 0.1736E-01\\
 $P32$  &   3   &   2.8353161E-17  &0.1875E-01\\
 \hline
\end{tabular}
\end{center}

\begin{center}
\begin{tabular}{ |p{2cm}|p{2.5cm}|p{5.5cm}|p{5.75cm}|  }
 \hline
 \multicolumn{4}{|c|}{RbH$_9$ structures} \\
 \hline
 Symmetry& Formula units & DOS($E_F$)(states/eV/f.u.) at 50GPa & DOS($E_F$)(states/eV/f.u.) at 150GPa\\
 \hline
 $C2$   &   3   &   2.4139685E-11   &0.1720E-03\\
 $C2/c$ &   2   &   0.2987E-76  &0.1975E-01\\
 $C2/m$ &   1   &   2.129423E-11    &0.2724\\
 $Cc$   &   2   &   0.4171E-07  &0.2632E-01\\
 $Cccm$ &   2   &   0.7734E-40  &0.6590E-01\\
 $Cm$   &   2   &   1.759505E-08    &0.2133E-01\\
 $Cmcm$ &   2   &   3.0176185E-12   &0.2983\\
 $Cmmm$ &   2   &   0.3413E-04  &0.3413E-04\\
 $Ima2$ &   2   &   5.0152385E-21   &0.6977E-09\\
 $P2_1/m$   &   2   &   0.1489E-75  &0.042855\\
 $P6_3/mmc$ &   2   &   0.2508E-08  &0.2574E-01\\
 $P6_422$   &   3   &   0.9160E-61  &1.38955E-05\\
 $Pmna$ &   2   &   0.6679E-10  &0.3645\\
 $R\bar{3}$ &   3   &   0.9480E-26  &0.1749\\
 \hline
\end{tabular}
\end{center}

\begin{center}
\begin{tabular}{ |p{2cm}|p{2.5cm}|p{5.5cm}|p{5.75cm}|  }
 \hline
 \multicolumn{4}{|c|}{RbH$_{11}$ structures} \\
 \hline
 Symmetry& Formula units & DOS($E_F$)(states/eV/f.u.) at 50GPa & DOS($E_F$)(states/eV/f.u.) at 150GPa\\
 \hline
 $Cmc2_1$   &   2   &   0.3969E-22  &0.1478\\
 $Imm2$ &   3   &   0.1788  &1.0960\\
 $Immm$ &   1   &   0.3101  &0.2932\\
 $Immm$ &   2   &   0.3102  &0.2911\\
 $Pmmn$ &   2   &   0.6278E-13  &0.5039\\
 \hline
\end{tabular}
\end{center}

\begin{center}
\begin{tabular}{ |p{2cm}|p{2.5cm}|p{5.5cm}|p{5.75cm}|  }
 \hline
 \multicolumn{4}{|c|}{RbH$_{12}$ structures} \\
 \hline
 Symmetry& Formula units & DOS($E_F$)(states/eV/f.u.) at 50GPa & DOS($E_F$)(states/eV/f.u.) at 150GPa\\
 \hline
 $C2$   &   2   &   0.91805 &0.7946\\
 $C2/c$ &   2   &       1.153   &0.95985\\
 $C2/m$ &   1   &   0.5849  &0.4330\\
 $Cmcm$ &   2   &   0.90505 &0.6612\\
 $Fmmm$ &   1   &   0.5736  &0.4078\\
 $Immm$ &   1   &   0.5564  &0.3666\\
 $Immm$ &   2   &   0.5515  &0.3664\\
 $P2/c$ &   2   &   0.8955  &0.8697\\
 $R\bar{3}m$    &   2   &   0.6024  &0.40845\\
 $P6_3/mmc$ &   2   &   --  &0.7904\\
 $Cmc2_1$   &   2   &   --  &1.00595\\
 \hline
\end{tabular}
\end{center}

\begin{figure}[H]
    \centering
    \includegraphics[width=\textwidth]{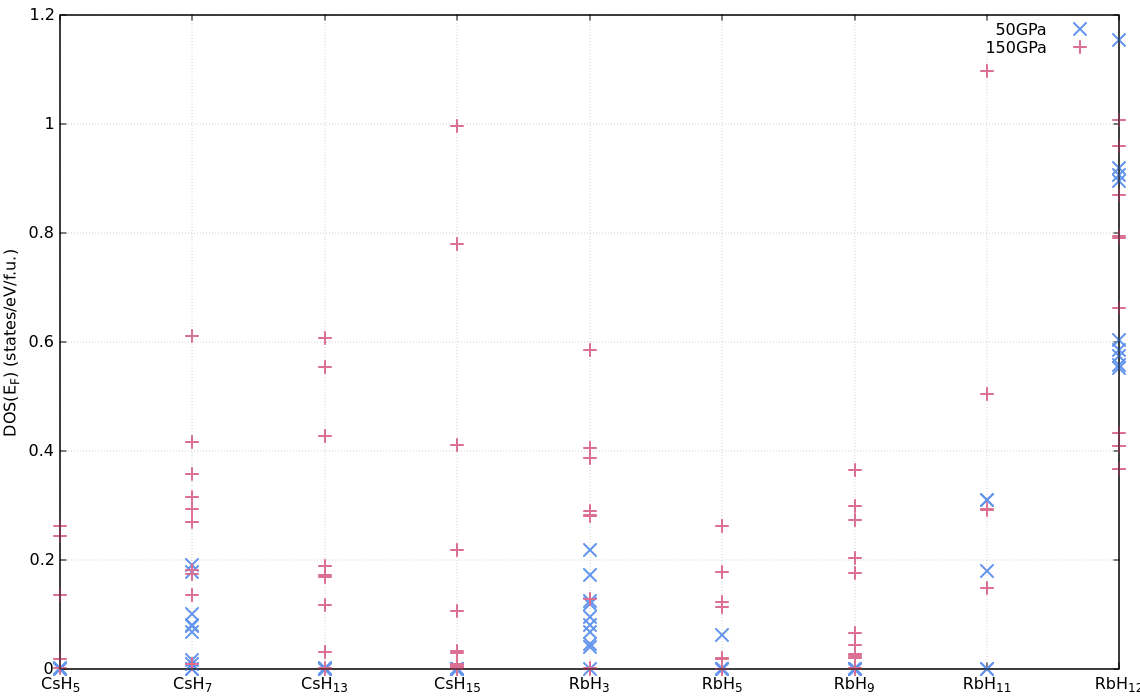}
    \caption{A plot summarising DOS($E_F$) at 50 and 150\ GPa for structures found using AIRSS.}
    \label{fig:dos_summary}
\end{figure}

\end{widetext}

\section{Testing further screening methods}
In this section we compare $\eta_j$ values calculated using Gaspari-Gyorffy, (GG) theory \cite{gaspari1972} and hydrogen DOS ratios ($N_H(E_F)/N(E_F)$) for various hydrides, for calculating $T_c$s to assess potential screening methods. The preliminary test cases chosen are various structures of LaH$_{10}$ and YH$_{10}$, so as to compare to Ref.\ \cite{shipley2020} and $I4/mmm$-FeH$_5$, so as to compare to the work of Ref.\ \cite{kvashnin2018} which reported superconductivity in this system and Ref.\ \cite{heil2018} which disputed this claim.

\begin{widetext}
\begin{center}
\begin{tabular}{ |p{3cm}|p{2.5cm}|p{3cm}|p{2.5cm}|p{2.5cm}|p{3cm}|  }
 \hline
 \multicolumn{6}{|c|}{Test cases} \\
 \hline
 Structure& Pressure/GPa &Calculated $T_c$ (K) & $\eta_H$ (eV/$\AA^2$) & $\eta_{X}$(eV/$\AA^2$) & $N_H(E_F)/N(E_F)$\\
 \hline
 $Fm\bar{3}m$-LaH$_{10}$ &250 &   234-259   & \cite{shipley2020}  5.58906  &0.50931 &0.28996 \\
 $P6_3/mmc$-LaH$_{10}$ &250 &  224-250   & \cite{shipley2020}  5.45099  &0.57450& 0.29676 \\
 $C2/m$-LaH$_{10}$  &250 &  205-228   & \cite{shipley2020}  4.48316  &0.46706& 0.28758\\
 $Fm\bar{3}m$-YH$_{10}$ &400 &   247-282 \cite{shipley2020}  &   11.45718 &2.11339& 0.32971\\
 $Cmcm$-YH$_{10}$ &400 &  233-270 \cite{shipley2020}  &   10.49255  &1.93803& 0.29863\\
 $P6_3/mmc$-YH$_{10}$  &400 &  223-262 \cite{shipley2020}   &   9.85728  &1.63951& 0.30800\\
 %$Im\bar{3}m$-YH$_6$ &160 & 223-247 \cite{shipley2020}   &   4.44873  &1.63826& 0.34787\\
 $I4/mmm$-FeH$_5$ &  150& 33.6-45.8 \cite{kvashnin2018}, $\leq$1 \cite{heil2018}   &   0.05468  &0.46240& 0.04091\\
 %$I4/mmm$-NdH$_4$ & 100 & 31-51 \cite{zhou2019neodymium}   &   0.09756  &1.05711& 0.01437\\
 %$C2/c$-NdH$_7$ & 100 & $\sim$0.01 \cite{zhou2019neodymium}   &   0.16644  &0.09941& 0.01006\\
 %$P6_3/mmc$-NdH$_9$ & 120 &  15-25 \cite{zhou2019neodymium}   &   0.16994  &0.03887& 0.01995\\
 %$Im\bar{3}m$-CaH$_6$ & 150 &   220-235 \cite{wang2012}   &   2.67506  &0.37970& 0.51126\\
 %$Fd\bar{3}m$-Li$_2$MgH$_{16}$ & 250 & 473 \cite{sun2019} & 5.02036 & 0.14702& 0.33024\\
 %$Im\bar{3}m$-H$_3$S & 200 & 191-204 \cite{duan2014} & 5.21134 & 4.24877& 0.25999\\
 %$Cmcm$-AsH & 350 & 20.2-21.2 \cite{fu2016} & 2.19678 & 3.42687& 0.06730\\
 %$C2/c$-AsH$_8$ & 350 & 141 \cite{fu2016} & 4.51675 & 1.15555& 0.17811\\
% $Pnma$-SbH & 300 & $\sim$6.8 \cite{fu2016} & 0.68405 & 43.1372& 0.03964\\
 %$Pmmn$-SbH$_3$ & 300 & 25.9 \cite{fu2016} & 3.91360 & 1.68235& 0.11186\\
 %$P6_3/mmc$-SbH$_4$ & 300 & 93.9 \cite{fu2016} & 3.69085 &1.49097& 0.11451\\
 \hline
\end{tabular}
\end{center}
\end{widetext}
In addition to calculating the values in the Table, comparing our $\eta_j$ values for H$_3$X (X=S, As, Se, Br, Sb, Te and I) to those in Ref.\ \cite{chang2019} allowed a rough validation of our implementation of GG theory against the NRL version. An extremely low $\eta_H$ is calculated for FeH$_5$, indicating that this material will not be a good superconductor - this result reflects the calculations of Ref.\ \cite{heil2018}.
\newline
\newline The average phonon frequencies for different structures are often very similar when we are considering the same stoichiometry at the same pressure. If we assume that the average phonon frequencies in such cases are exactly equivalent, we arrive at a potential way of cheaply estimating the $T_c$ \textit{order} between structures. Using our results to rank the various LaH$_{10}$ and YH$_{10}$ structures (considering $\eta_H$ only since this term tends to have the most impact on the total $\lambda$), we arrive at $Fm\bar{3}m>P6_3/mmc>C2/m$ for LaH$_{10}$ and $Fm\bar{3}m>Cmcm>P6_3/mmc$ for YH$_{10}$. Agreement with the calculated $T_c$ order is seen in both cases (while the $N_H(E_F)/N(E_F)$ predicts a slightly incorrect ordering for LaH$_{10}$).

\section{Selecting promising candidates}
Based on the initial DOS calculations, RbH$_{12}$ was chosen as one of the most promising compositions as it has several metallic structures down to 50\ GPa. Looking at the RbH$_{12}$ enthalpy plot, the lowest enthalpy structures across the whole pressure range are $Cmcm$ (2 formula units), $Cmcm$ (3 formula units), $C2/m$ (1 formula unit), $Immm$ (1 formula unit) and $Immm$ (2 formula units). On refining the structures, $Cmcm$-3 reduced to the structure with 2 formula units; similar behaviour is seen for the two $Immm$ structures. All of these structures are metallic at 50\ GPa according to the initial DOS calculations. We calculate the $N_H(E_F)/N(E_F)$ ratios and $\eta_H$ values for these structures at 50, 100 and 150\ GPa in order to test these screening methods - these values are displayed in the Table below. $C2/m$-1 and $Immm$-1 have similar (and fairly high) DOS ratios and $\eta_H$ even at these low pressures. $Cmcm$-2 also appears promising. We perform full electron-phonon superconductivity calculations on these 3 structures at various pressures.

According to the initial DOS calculations, CsH$_7$ is also a promising stoichiometry. The CsH$_7$ enthalpy plot shows a group of structures that are competitive at $P<90$\ GPa ($Imm2$-2, $I4mmm$-2, $Cm$-3, $P4mm$-2) and then another competitive group at slightly higher pressures ($I4mm$-1, $P1$-3, $Cmc2_1$-2, $I4mm$-2). The Table below shows DOS ratios and $\eta_H$ values for these structures at 100\ GPa. Many of the structures look fairly promising and we proceed with running full electron-phonon superconductivity calculations. The test screening methods described here suggest that $Cmc2_1$-2, $I4mm$-1 and $P1$-3 may give the highest $T_c$ at a given pressure.

We also consider CsH$_{15}$. None of the structures are particularly metallic at 50\ GPa according to initial DOS calculations, with a few becoming metallic at 150\ GPa - these include $Ima2$-2, $P2_1$-2, $P2/m$-2, $Pm$-2 and $Cmc2_1$-2. Looking at the enthalpy plot, the most competitive structures are $Cm$-1, $Cm$-2, $P1$-1, $P1$-2, $Pmn2_1$-2, $Imm2$-1, $Pm$-2, $P2_1m$-2, $P2_1$-2 and $Cmc2_1$-2. These results show that $Pm$-2, $Cmc2_1$-2, $P2_1m$-2 and $P2_1$-2 are the only structures which are both energetically competitive and metallic. We calculate the DOS ratio and scattering theory parameters for some of the competitive structures at 150\ GPa, but do not go on to perform electron-phonon calculations.

We also consider RbH$_3$. We first refine the set of structures we will look at by considering the enthalpy plots. Ignoring the most energetically unfavourable structures at the static-lattice level (here, the highest 3) and noting that $Cmmm$-2 and $Cmmm$-1 correspond to the same structure, we are left with $I4_1/amd$-2, $C2/c$-2, $Cmmm$-1, $Cccm$-2, $P2/m$-3, $P6_222$-3 and $Pmma$-2. However, $I4_1/amd$-2 and $C2/c$-2 were not sufficiently metallic at 100\ GPa to be able to calculate $\eta$ values or DOS ratios; this reflects the results shown in the initial DOS Table and explains why they are not included in the Table below. We performed electron-phonon calculations for two of the predicted best structures of RbH$_3$, but found their superconducting temperatures to be extremely low; this can be explained by their layered nature as addressed in the main text. We have not considered the other structures any further.

We also consider RbH$_{11}$ - for this stoichiometry, $Cmc2_1$-2 is most stable at lower pressures and $Pmmn$-2 and $Immm$-1 are competitive at higher pressures. $Cmc2_1$ and $Pmmn$ were not metallic enough at the pressure considered in order to be included in the Table. We did not perform electron-phonon calculations for any RbH$_{11}$ structures - $Immm$ could be promising, but since the enthalpy plot suggests it becomes competitive only above $\sim$ 125\ GPa, we have not considered it further.

As addressed in the main text, our calculated $T_c$ values allow us to further assess the screening/ranking methods tested in this work. The $\eta_H$ values correctly predict $T_c$ ordering for the RbH$_{12}$ structures at fixed pressure and $N_H(E_F)/N(E_F)$ also comes close to doing so. $N_H(E_F)/N(E_F)$ appears to be much less predictive for the CsH$_7$ structures and the performance of $\eta_H$ is also mixed. We note that in some cases the trends predicted by the DOS ratio and the scattering calculation disagree considerably (see $I4/mmm$-CsH$_7$ and $P1$-CsH$_7$). This can occur for cases in which the structure has a low total DOS at $E_F$ and the DOS ratio is therefore artificially inflated.

\begin{widetext}
\begin{center}
\begin{tabular}{ |p{3cm}|p{3cm}|p{2.5cm}|p{2.5cm}|p{3cm}|p{2cm}|  }
 \hline
 \multicolumn{6}{|c|}{Candidate structures} \\
 \hline
 Structure& Pressure/GPa & $\eta_H$ (eV/$\AA^2$) & $\eta_{X}$(eV/$\AA^2$) & $N_H(E_F)/N(E_F)$ & $T_c$/K\\
 \hline
 $C2/m$-RbH$_{12}$ & 50 & 0.90524 & 0.04356& 0.15568 & 108 \\
 $C2/m$-RbH$_{12}$ & 100 &1.45034 & 0.09396& 0.18652 & 129 \\
 $C2/m$-RbH$_{12}$ & 150 &1.97268 & 0.15800& 0.20542 & 133 \\
 %$Cmcm$-RbH$_{12}$ &50  &  --  &-- &-- & -- \\
 $Cmcm$-RbH$_{12}$ &100  &   1.21876  &0.05165 &0.17611 & 82 \\
 %$Cmcm$-RbH$_{12}$ &150  &   --  &-- &-- & -- \\
 $Immm$-RbH$_{12}$ & 50 & -- &-- &-- & 115\\
 $Immm$-RbH$_{12}$ & 100 & 1.45016 &0.09344 &0.18709 & 119\\
 $Immm$-RbH$_{12}$ & 150 & -- &-- &-- & 126\\
 $Cmc2_1$-CsH$_7$ & 100 & 0.78707 & 0.03099 &0.15403 & 89\\
 $I4mm$-CsH$_7$ & 100 & 0.78877 & 0.02803 & 0.15737 & 34\\
 $I4/mmm$-CsH$_7$ & 100 & 0.00753 & 0.00262 & 0.19421 & 10\\
 $Imm2$-CsH$_7$ & 100 & 0.02101 & 0.00217 & 0.20200 & --\\
 $P1$-CsH$_7$ & 100 & 0.77077 & 0.02638& 0.15734 & 90\\
 $P4mm$-CsH$_7$ & 100 & 0.53524 & 0.01394 & 0.21268 & 33\\
 $Cm$-CsH$_7$ & 100 & 0.57309 & 0.01343 & 0.21174 & 5\\
 $Pm$-CsH$_{15}$ & 150 & 0.95176 & 0.03376 & 0.19481 & --\\
 $P2_1m$-CsH$_{15}$ & 150 & 0.94744 & 0.03425 & 0.19679 & --\\
 %$Cmc2_1$-CsH$_{15}$ & 150 & -- & -- & -- & --\\
 %$P2_1$-CsH$_{15}$ & 150 & -- & -- & -- & --\\
 %$I4_1/amd$-RbH$_3$	& 100 & -- & --& -- & --\\
 %$C2/c$-RbH$_3$ & 100 & -- & -- & -- & --\\
 $Cccm$-RbH$_3$ & 100 & 0.22802 & 0.07998 & 0.10363 & --\\
 $Cmmm$-RbH$_3$ & 100 & 0.24435 & 0.06429 & 0.14831 & 0\\
 $P2/m$-RbH$_3$ & 100 & 0.22855 & 0.06377 & 0.11610 & --\\
 $P6_222$-RbH$_3$ & 100 & 0.23745 & 0.05618 & 0.11044 & --\\
 $Pmma$-RbH$_3$ & 100 & 0.28952 & 0.06610 & 0.11262 & 0\\
 %$Cmc2_1$-RbH$_{11}$ & 100 & -- & -- & -- & --\\
 %$Pmmn$-RbH$_{11}$ & 100 & -- & -- & -- & --\\
 %$Imm2$-RbH$_{11}$ & 100 & -- & -- & -- & --\\
 $Immm$-RbH$_{11}$ & 100 & 1.55705 & 0.03130 & 0.23464 & --\\
 \hline
\end{tabular}
\end{center}
\end{widetext}

\section{Electron-Phonon coupling in DFT}
Typically, within DFT the nuclear coordinates, $R$, are treated as fixed and the electronic Kohn-Sham system is solved within the fixed nuclear potential. In order to calculate the effects of electron-phonon coupling from within the DFT formalism we must consider leading-order corrections to the Born-Oppenheimer approximation in nuclear displacements. Expanding our Kohn-Sham potential in terms of these displacements leads to
\begin{equation}
\label{eq:ks_pertubation}
    V_{KS}(R+\delta R) = V_{KS}(R) + \sum_{\kappa, p} \frac{\partial V_{KS}}{\partial R_{\kappa, p}} \cdot \delta R_{\kappa, p} + O(\delta R^2).
\end{equation}
where $R_{\kappa, p}$ is the position of atom $\kappa$ in unit cell $p$. An atomic displacement of an atom can be written in terms of phonon creation and annihilation operators \cite{DFT_ELEC_PHONON} as
\begin{equation}
    \delta R_{\kappa, p} = \frac{1}{\sqrt{N_pM_\kappa}} \sum_{q\nu} e^{iq\cdot R_p} \frac{1}{\sqrt{2\omega_{q\nu}}} \left( a_{q\nu}+a_{-q\nu}^\dagger \right) e_{\kappa\nu}(q)
\end{equation}
where $e_{k\nu}(q)$ and $\omega_{q,\nu}$ are, respectively, the eigenvector and frequency of the phonon mode with creation operator $a_{q\nu}^\dagger$. $R_p$ is the position of the $p$\textsuperscript{th} unit cell within the periodic cell, of which there are $N_p$. $M_\kappa$ is the mass of atom $\kappa$. Substituting this into Eq.\ \ref{eq:ks_pertubation} we obtain
\begin{equation}
    V_{KS}(R + \delta R) = V_{KS}(R) + \frac{1}{\sqrt{N_p}} \sum_{q\nu} G_{q\nu} (a_{q\nu} + a_{-q\nu}^\dagger)
\end{equation}
where
\begin{equation}
    G_{q\nu} = \frac{1}{\sqrt{2\omega_{q\nu}}} \sum_\kappa \frac{e_{\kappa\nu}(q)}{\sqrt{M_\kappa}} \cdot \sum_p e^{iq \cdot R_p} \frac{\partial V_{KS}}{\partial R_{\kappa,p}}
\end{equation}
This allows us to write down the resulting electron-phonon coupling Hamiltonian in second-quantized form as 
\begin{equation}
\begin{aligned}
    &H_{ep} (\delta R) \\&= \sum_{nkn'k'} \bra{n,k} V_{KS}(R + \delta R) - V_{KS}(R) \ket{n', k'} c_{nk}^\dagger c_{n'k'} \\
    &= \frac{1}{\sqrt{N_p}} \sum_{q\nu} \left[\sum_{nkn'k'} \bra{n,k} G_{q\nu} \ket{n'k'} c_{nk}^\dagger c_{n',k'}\right] (a_{q\nu} + a_{-q\nu}^\dagger)
\end{aligned}
\end{equation}
where $c_{nk}^\dagger$ creates a Kohn-Sham electron in orbital $n$, wavevector $k$ (i.e., occupies the Bloch state $u_{nk}(x)\exp(ik\cdot x)/\sqrt{N_p}$). Substituting our definition of $G_{q\nu}$ we have
\begin{equation}
\label{eq:big_g_matrix_elements}
\begin{aligned}
    &\bra{n,k} G_{q\nu} \ket{n'k'} \\&= \frac{1}{\sqrt{2\omega_{q\nu}}} \sum_\kappa \frac{e_{\kappa\nu}(q)}{\sqrt{M_\kappa}} \cdot \sum_p e^{iq \cdot R_p} \bra{n,k} \frac{\partial V_{KS}}{\partial R_{\kappa,p}} \ket{n',k'}
\end{aligned}
\end{equation}
Now
\begin{equation}
\begin{aligned}
    &\bra{n,k} \frac{\partial V_{KS}}{\partial R_{\kappa,p}} \ket{n',k'} \\&= \int N_p^{-1/2} u_{nk}^*(x)e^{-ik\cdot x} \frac{\partial V_{KS}}{\partial R_{\kappa,p}}(x) N_p^{-1/2} u_{n'k'}(x)e^{ik'\cdot x} \;dx \\
    &= \int N_p^{-1/2} u_{nk}^*(x-R_p)e^{-ik\cdot (x-R_p)} \frac{\partial V_{KS}}{\partial R_{\kappa,p}}(x-R_p) \\&\hspace{0.5cm} \times N_p^{-1/2} u_{n'k'}(x-R_p)e^{ik'\cdot (x-R_p)} \;dx \\
    &= e^{iR_p\cdot(k-k')} \int_{\text{1\textsuperscript{st} unit-cell}} \hspace{-1cm} u_{nk}^*(x)e^{-ik\cdot x} \frac{\partial V_{KS}}{\partial R_{\kappa,0}}(x) u_{n'k'}(x)e^{ik'\cdot x} \;dx
\end{aligned}
\end{equation}
where in the last line we have used Bloch's theorem and the fact that
\begin{equation}
    \frac{\partial V_{KS}}{\partial R_{\kappa,p}}(x - R_p) = \frac{\partial V_{KS}}{\partial R_{\kappa,0}}(x)
\end{equation}
where $R_{\kappa,0}$ is the position of atom $\kappa$ in the first unit cell. We may now write Eq.\ \ref{eq:big_g_matrix_elements} as
\begin{equation}
\begin{aligned}
    &\bra{n,k} G_{q\nu} \ket{n'k'} \\&= \frac{1}{\sqrt{2\omega_{q\nu}}} \sum_\kappa \frac{e_{\kappa\nu}(q)}{\sqrt{M_\kappa}} \cdot \bra{n,k} \frac{\partial V_{KS}}{\partial R_{\kappa,0}} \ket{n',k'}_{\text{uc}} \underbrace{\sum_p e^{i(q + (k-k')) \cdot R_p}}_{N_p\delta_{q,k-k'}}
\end{aligned}
\end{equation}
where the subscript ``uc" on the ket means integration only over the first unit cell. Finally we obtain the DFT electron-phonon coupling Hamiltonian
\begin{equation}
\begin{aligned}
    H_{ep} = \frac{1}{\sqrt{N_p}} \sum_{q\nu knm} \bra{m,k+q} G_{q\nu,\text{uc}} \ket{n,k}_{\text{uc}} \\ \times c_{m,k+q}^\dagger c_{n,k} (a_{q\nu} + a_{-q\nu}^\dagger)
\end{aligned}
\end{equation}
where we have defined
\begin{equation}
    G_{q\nu,\text{uc}} = \frac{1}{\sqrt{2\omega_{q\nu}}} \sum_\kappa \frac{e_{\kappa\nu}(q)}{\sqrt{M_\kappa}} \cdot \frac{\partial V_{KS}}{\partial R_{\kappa,0}}
\end{equation}
This allows us to write down the Hamiltonian for an interacting Kohn-Sham-electron-phonon system, correct to first order in electron-phonon coupling constants $g_{mn\nu}(k,q) = \bra{m,k+q} G_{q\nu,\text{uc}} \ket{n,k}_{\text{uc}}$:
\begin{equation}
\label{eq:electron_phonon_hamiltonian}
\begin{aligned}
    H &= \underbrace{\sum_{kn} \epsilon_{nk} c_{nk}^\dagger c_{nk}}_{\text{Electronic dispersion}} + \underbrace{\sum_{q\nu} \omega_{q\nu} \left(a_{q\nu}^\dagger a_{q\nu} + \frac{1}{2}\right)}_{\text{phonon dispersion}} +\\
    &\underbrace{\frac{1}{\sqrt{N_p}} \sum_{kqmn\nu} g_{mn\nu}(k,q)c_{m,k+q}^\dagger c_{nk} \left( a_{q\nu} + a_{-q\nu}^\dagger\right).}_{\text{electron-phonon coupling}}
\end{aligned}
\end{equation}
From the parameters in this Hamiltonian we can also define the electron-phonon coupling strength associated with each phonon mode, $\lambda_{q\nu}$, and the isotropic Eliashberg spectral function, $\alpha^2F(\omega)$
\begin{equation}
\begin{aligned}
    \lambda_{q,\nu} &= \frac{1}{N(\epsilon_F)\omega_{q\nu}\Omega_{\text{BZ}}} \\&\times \sum_{nm} \int_{\text{BZ}} |g_{mn\nu}(k,q)|^2 \delta(\epsilon_{n,k} - \epsilon_F)\delta(\epsilon_{m,k+q} - \epsilon_F) dk
\end{aligned}
\end{equation}
\begin{equation}
    \alpha^2F(\omega) = \frac{1}{2\Omega_{\text{BZ}}}\sum_\nu \int_{\text{BZ}} \omega_{q\nu}\lambda_{q\nu}\delta(\omega - \omega_{q\nu}) dq
\end{equation}
from which we may approximate the critical temperature using the McMillan formula \cite{mcmillan_formula}
\begin{equation}
    T_c = \frac{\omega_{\text{log}}}{1.2} \exp\left(\frac{-1.04(1+\lambda)}{\lambda(1-0.62\mu^*) - \mu^*}\right)
\end{equation}
where
\begin{equation}
    \lambda = \sum_{q\nu} \lambda_{q\nu} = 2\int \alpha^2F(\omega) \frac{d\omega}{\omega},
\end{equation}
\begin{equation}
    \omega_{\text{log}} = \exp\left(\frac{2}{\lambda}\int \alpha^2F(\omega)\log(\omega) \frac{d\omega}{\omega} \right)
\end{equation}
and $\mu^*$ is the Morel-Anderson pseudopotential \cite{mu_star}, which is typically treated as an empirical parameter with values between 0.1 and 0.2.

\begin{figure}
\begin{subfigure}
    \centering
    \includegraphics[width=\columnwidth]{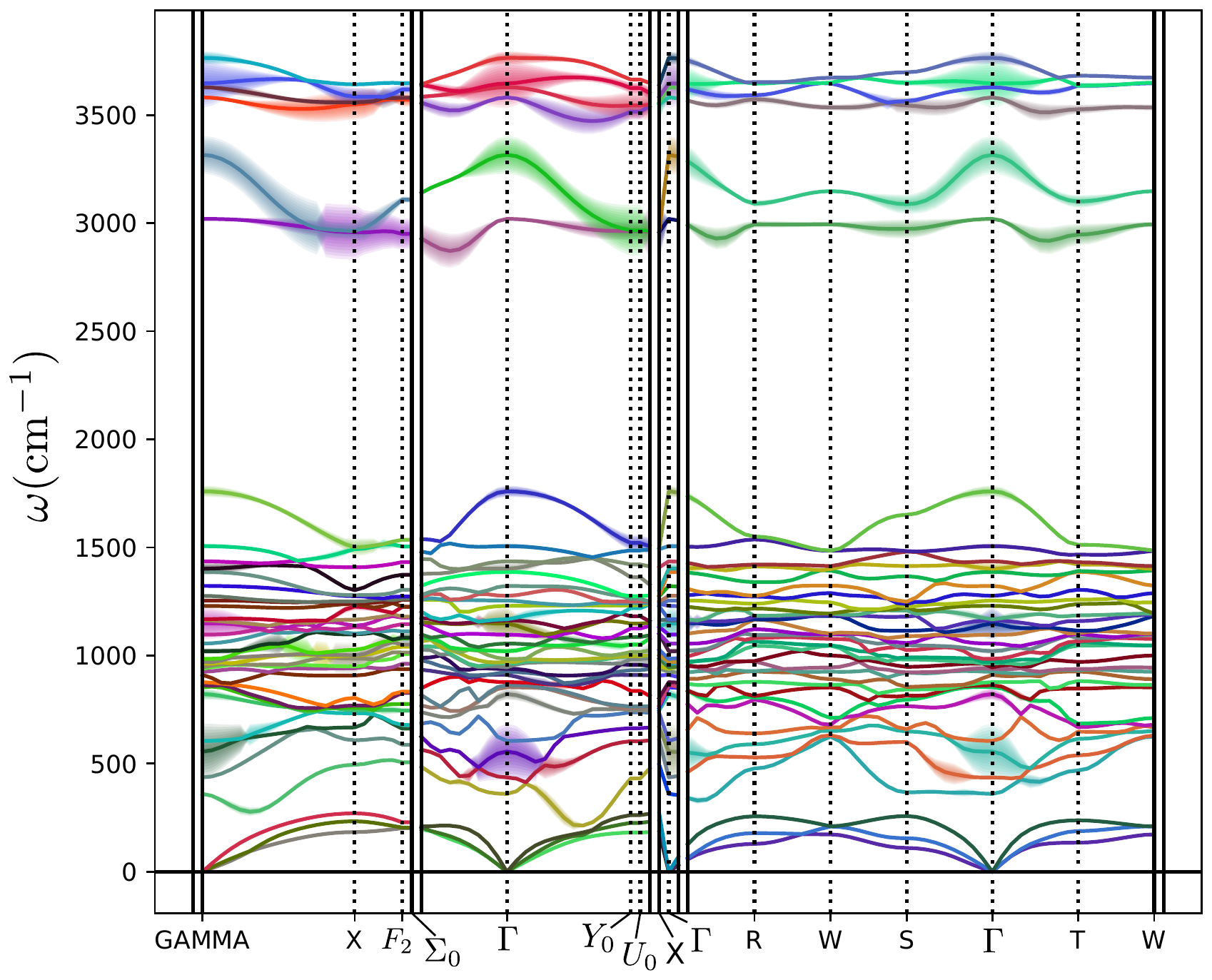}
    \caption{Phonon dispersion of the $Immm$ phase of RbH$_{12}$ showing linewidths due to electron-phonon coupling, calculated using DFPT. Note that the high-frequency phonon modes due to vibrations of the hydrogen cage have large linewidths, resulting in a high critical temperature.}
    \label{fig:rbh12_imma_phonons}
\end{subfigure}

\begin{subfigure}
    \centering
    \includegraphics[width=\columnwidth]{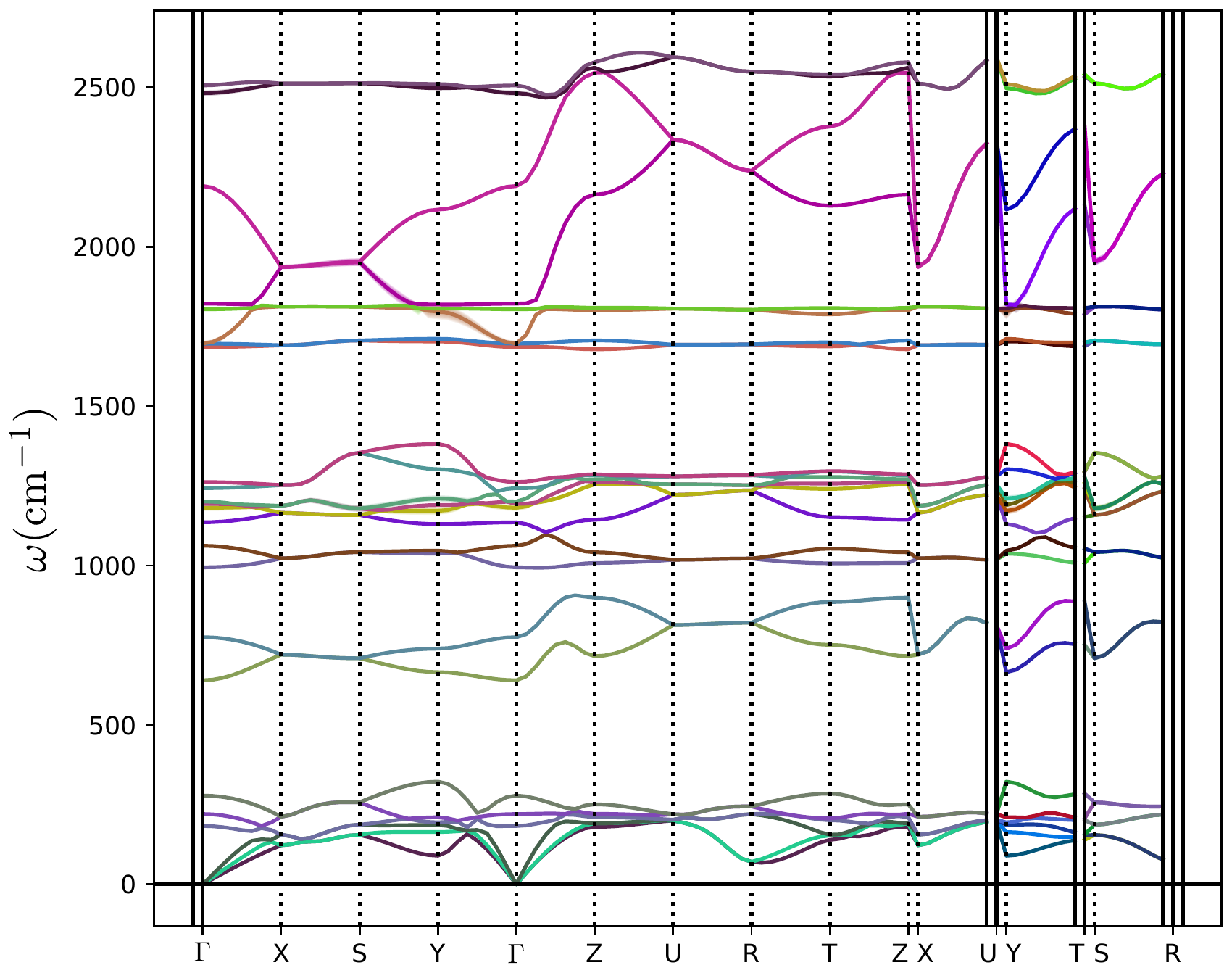}
    \caption{Phonon dispersion of the $Pmma$ phase of RbH$_{3}$ showing linewidths due to electron-phonon coupling, calculated using DFPT. Because of the layered structure, there are no high-frequency phonon modes from hydrogen-only vibrations (in contrast to the case in Fig.\ \ref{fig:rbh12_imma_phonons}).}
    \label{fig:rbh3_pmma_phonons}
\end{subfigure}
\end{figure}

\bibliography{references.bib}